\documentclass[3p]{elsarticle}

\biboptions{sort&compress}

\journal{Strain}

\usepackage{amsmath, amsthm, amssymb, amsfonts}
\usepackage[table,dvipsnames]{xcolor}
\usepackage{mathtools}
\usepackage{soul}
\usepackage{setspace} 
\usepackage{float}

\usepackage{threeparttable}
\usepackage{etoolbox}
\usepackage{longtable}
\usepackage{multirow}
\usepackage[colorinlistoftodos,prependcaption,textsize=tiny,textwidth=3.5cm]{todonotes}
\usepackage{booktabs}
\usepackage[utf8]{inputenc}
\usepackage{enumitem}
\usepackage{epsfig}
\usepackage{hyperref}

\usepackage{optidef}

\usepackage{tabularx} 
\usepackage{tabu} 
\usepackage{booktabs} 
\usepackage{multirow}
\usepackage{adjustbox}

\usepackage[font=small,labelfont=bf]{caption} 
\usepackage[font=small,labelfont=bf]{subcaption} 

\definecolor{darkgreen}{rgb}{0.1,0.8,0.1}

\newcommand{\degree}{$^{\circ}$}
\newcommand{\psslip}{\textbf{+}SSLIP }

\newcommand{\us}{\textsubscript}

\begin{document}
\singlespace

\begin{frontmatter}

\title{\textbf{+}SSLIP: automated Radon-assisted and rotation-corrected identification of complex HCP slip system activity fields from DIC Data}

\author[mymainaddress,secondAddress]{T. Vermeij \fnref{fn1}}\address[mymainaddress]{Dept. of Mechanical Engineering, Eindhoven University of Technology, 5600MB Eindhoven, The Netherlands}
\address[secondAddress]{Laboratory for Mechanics of Materials and Nanostructures, Swiss Federal Laboratories for Materials Science and Technology (EMPA), Feuerwerkerstrasse
39, 3602 Thun, Switzerland}
\author[mymainaddress]{G. Slokker \fnref{fn1}}
\author[mymainaddress]{C.J.A. Mornout}
\author[mymainaddress]{D. König}
\author[mymainaddress]{J.P.M. Hoefnagels*}
\cortext[mycorrespondingauthor]{Corresponding author}
\fntext[fn1]{These authors contributed equally.}
\ead{j.p.m.hoefnagels@tue.nl}

\begin{abstract}
Identification of crystallographic slip in metals and alloys is crucial to understand and improve their mechanical behavior. Recently, a novel slip system identification framework, termed SSLIP (for Slip Systems based Local Identification of Plasticity), was introduced to leap from conventional trace-based identification to automated, point-by-point identification, exploiting the full deformation kinematics. Using sub-micron-scale Digital Image Correlation (DIC) deformation fields aligned to Electron Backscatter Diffraction (EBSD) data, SSLIP matches the measured in-plane displacement gradient tensor to the kinematics of the optimal combination of multiple slip system activities, at each DIC datapoint. SSLIP was demonstrated to be successful on virtual and experimental case studies of FCC and BCC metals. However, for more challenging HCP crystal structures, the complete identification of all slip systems was found to be more challenging, posing limitations on automation and flexibility. To extend the capabilities of SSLIP, we propose an extended framework, hereinafter referred to as the \psslip method, which includes (i) a preselection of slip systems using a Radon transform, (ii) robustness to measured rigid body rotation by simultaneous identification of the local rotation field, (iii) identification of the two best matching slip systems for each data point, and (iv) a procedure to determine groups of slip systems with in-plane displacement gradient tensors that cannot be discriminated. This procedure yields the full (HCP) slip system activity maps for every slip system in each grain. The resulting objective identification method does not rely on the Schmid factor to select which slip system is active at each point. We show how slip systems from multiple slip families are successfully identified on virtual and real experiments on a Zn polycrystalline coating. \url{https://doi.org/10.1111/str.70000}

\end{abstract}


\end{frontmatter}

\section{Introduction}

Understanding the mechanisms of plastic deformation in metals and alloys, governed predominantly by crystallographic slip, is fundamental to advancing materials science and engineering. Crystallographic slip, comprising the movement of dislocations across slip planes in specific crystallographic slip directions, is a crucial plastic deformation mechanism that governs how metals respond to external stresses. The combination of a certain slip plane and slip direction forms a slip system and is used to describe the kinematics and geometry of plastic slip. However, accurately identifying and quantifying slip system activities in experiments poses significant fundamental challenges, that are exacerbated by various complex phenomena, such as the complex interplay of multiple slip systems and the presence of diffuse or intersecting slip bands, which are hard to measure with existing characterization techniques \cite{lunt2018quantification, githens2020characterizing, harte2020effect, vermeij2022nanomechanical, vermeij2023damage, vermeij2023testing2d}.

Over the years, various slip system identification methods have been proposed and applied. Through Scanning Electron Microscopy (SEM), discrete and straight slip bands can be observed and matched to theoretical slip plane traces, based on measured crystal orientations through Electron Backscatter Diffraction (EBSD) \cite{Parisot2004DeformationModes, bridier2005analysis, Bieler009, echlin2016incipient}. Additionally, strain maps acquired through SEM-based Digital Image Correlation (SEM-DIC) can be employed for slip trace analysis \cite{OROZCOCABALLERO2017367, harr2021effect, Hu2023AutomatedAlloy, sperry2020slip}. However, in order to obtain a more complete identification, the slip direction also needs to be considered. Therefore, approaches such as the "Relative Displacement Ratio" (RDR) method \cite{chen_daly_2016} and the "Heaviside DIC method" \cite{BOURDIN2018307} aim to also extract the slip direction from the SEM-DIC displacement field. The Heaviside DIC method incorporates an additional step function within a DIC subset to deduce the slip trace and direction, assuming that slip is confined to a single slip plane resulting in a clean slip step \cite{BOURDIN2018307}. On the other hand, the RDR method analyzes regular DIC data by comparing displacement component ratios across a distinct slip trace to theoretical Burgers vector component ratios \cite{chen_daly_2016}. While these methods have proven effective in various studies \cite{XU2019376, STINVILLE2020172, sperry2021comparison, poole2024high, black2024micro}, their application is mostly limited to scenarios with well-defined slip traces, and they struggle with diffuse or complex slip, partly pertaining due to the resolution constraints of SEM-DIC.

To address the complexities of slip system identification and the limitations of current methodologies, Vermeij \textit{et al.} recently introduced an automated identification approach, termed as SSLIP (for Slip Systems based Local Identification of Plasticity) \cite{Vermeij2022AutomatedData}. This method leverages the detailed deformation data produced through DIC to facilitate a point-by-point identification of slip system activities across a sample surface. By solving an optimization problem at every SEM-DIC data point, SSLIP matches the measured in-plane kinematics - represented by displacement gradient tensor fields - to a combination of theoretical slip system kinematics derived from EBSD data after data alignment \cite{vermeij2022nanomechanical}. This novel approach not only enables the identification of slip system activity fields for each potential slip system, but also discloses the local variations in slip activity, particularly near grain and phase boundaries. Such detailed analysis holds significant promise for enhancing the understanding of plastic deformation mechanisms and allows for more reliable comparisons between experiments and theoretical simulations \cite{stinville2023insights, vermeij2023damage, vermeij2023testing2d, wijnen2024high}, as evidenced by the rapid adoption of the SSLIP method by the research community \cite{scherertensile, agius2023experimental, depriester2023crystal, yin2023three}.

However, as discussed by Vermeij \textit{et al.} \cite{Vermeij2022AutomatedData}, the automated and complete application of SSLIP on Hexagonal Close-Packed (HCP) materials is challenging due to its high anisotropy that enables a wide range of slip families. This results in a large number of slip systems with (very) similar in-plane kinematics, making them linearly dependent in the optimization problem \cite{Vermeij2022AutomatedData}. By analyzing virtual HCP experiments, it was found that in such demanding  cases, a preselection of slip systems to limit the number of possibly active slip systems was necessary to achieve a correct identification. While it was demonstrated  that  manual  preselection  of  slip  systems works for the virtual HCP experiment (without elastic rotations) and for single and bicrystal Zn experiments \cite{Konig2024TheCoating}, automation is  warranted in order  to  sustain  statistical investigations  of  slip  activities  in  multiple  grains. The issue is exacerbated by elastic rotations, which are not accounted for in the SSLIP optimization framework. Even though the apparent deformations due to elastic rotations are tiny, their effect in the presence of a large number of similar slip systems can often be large. As a result, a convincing, robust experimental demonstration of SSLIP on HCP is lacking, especially a demonstration of many HCP grains, for which automatic identification is crucial. 

Here we extend the current SSLIP method to make it a robust and automatic slip identification method for HCP (and other complex) crystals. Therefore, we first propose to implement an automated preselection step by means of a Radon transform \cite{helgason1999radon}, known for its ability to detect straight features, i.e. slip bands from strain maps \cite{Hu2023AutomatedAlloy}. We will demonstrate the efficiency of this automated preselection step in the presence of complex and diffuse slip. Subsequently, from the preselected slip systems, the best-fitting combination of two of the preselected systems will be identified for each individual data point separately. This step employs an optimization algorithm that not only seeks the optimal match based on the local kinematics but also incorporates a correction for any additional local rotation. This rotation correction is crucial as it accounts for rotational misalignment of the sample and grain rotations, that may otherwise impede accurate slip system identification, as explained above. By only allowing two slip systems per point, our approach enhances the robustness of the identification process, while avoiding problems in the iterative solution method due to multiple linear-dependent slip systems. We expect this to be a reasonable restriction, as it is unlikely that more than two slip systems are active in a single data point. Finally, we also identify groups of nondiscriminatory slip systems, of which the in-plane kinematics cannot be discerned. We term this extended methodology as \psslip to distinguish it from the original SSLIP method.

\begin{figure}[ht!]
    \centering
 \includegraphics[width=1.0\textwidth]{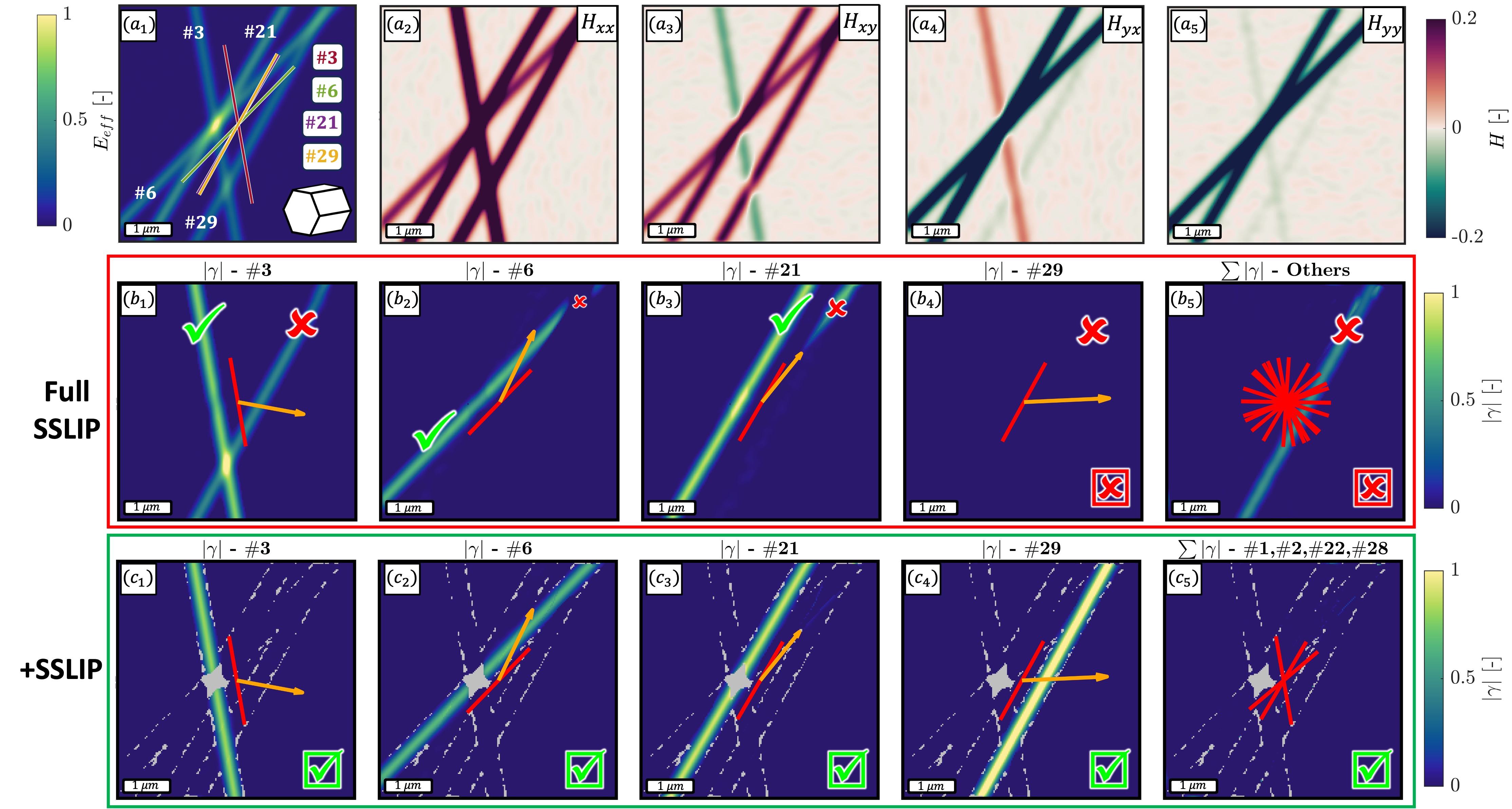}
    \captionsetup{width=1.0\linewidth} \caption{\textit{Virtual HCP SEM-DIC experiment, with (a$_1$) corresponding artificially generated effective strain field, $E_{eff}$, with strain bands marked and corresponding slip traces of the four included slip systems from different slip families, (a$_2$)-(a$_5$) the four in-plane displacement gradient tensor components, respectively $H^{exp}_{xx}$,$H^{exp}_{xy}$,$H^{exp}_{yx}$,$H^{exp}_{yy}$. (b) The identified slip activity fields using the original SSLIP method, reproduced from \cite{Vermeij2022AutomatedData}, with (b$_5$) showing the sum of all others slip activities that are not supposed to be active. (c) Identification result of the new \psslip approach. Green checkmarks indicate correctly identified slip systems, red crosses indicate erroneous ones.}}
    \label{fig:HCPvirtualexperiment}
\end{figure}

As a preview, we show the significant advancements made with \psslip on the challenging virtual HCP experiment presented in the original SSLIP paper (Section 7 in \cite{Vermeij2022AutomatedData}). \autoref{fig:HCPvirtualexperiment}(a) displays the strain field of this synthetic experiment, where four slip systems are active from four different slip families. The results from the original full SSLIP identification (without preselection), shown in \autoref{fig:HCPvirtualexperiment}(b), clearly show that several slip systems were misidentified, as marked by the red crosses. In contrast, \autoref{fig:HCPvirtualexperiment}(c) gives a preview of the results of the new \psslip approach, utilizing the Radon transform for automated preselection, showing a nearly perfect identification of all four slip systems, as discussed below.

The paper is structured as follows. In \autoref{sc:method}, the \psslip method is outlined in detail and demonstrated on the challenging virtual experiment that is introduced in \autoref{fig:HCPvirtualexperiment}. Next, in \autoref{sec:hcp_case_study}, an experimental SEM-DIC case study of HCP is used to demonstrate the strength of the \psslip method for selected HCP grains that show various complicated slip mechanisms, showing a robust identification of slip activity fields while considering no less than 30 slip systems from 5 different slip families. Finally, conclusions are summarized in \autoref{sec:concl}.

\section{Methodology}
\label{sc:method}

In this section, we outline the framework developed to extend the SSLIP methodology to \psslip, focusing on the identification of slip system activities within HCP materials. First, the virtual experiment and the original SSLIP method \cite{Vermeij2022AutomatedData} are reiterated, followed by the introduction of the new \psslip method. 

\subsection{Virtual Experiment and Reiteration of the Original SSLIP Method}
The virtual experiment, originally designed as the ultimate challenge for the SSLIP methodology \cite{Vermeij2022AutomatedData}, incorporates four overlapping slip systems from different slip families that represent the possible complexity of slip activity in an HCP material. The displacement field $\vec{u}$ is generated using the following slip systems: Basal-3 (\#3), PrismI-3 (\#6), PyrICA-9 (\#21), and PyrII-5 (\#29). As an example, "PrismI-3 (\#6)" refers to the third system of the first order Prismatic slip system family, which is the sixth slip system overall. \autoref{fig:HCPvirtualexperiment}(a) already showed the effective strain field $E_{\text{eff}}$, based on the displacement gradient tensor $\mathbf{H}^{exp}$, which is calculated by taking the gradient of the displacement field:
\begin{equation}
\mathbf{H}^{exp} = \vec{\nabla}_0 \vec{u} =
  \begin{bmatrix}
   H_{xx}^{exp} & H_{xy}^{exp} \\
   H_{yx}^{exp} & H_{yy}^{exp}
  \end{bmatrix}.
\label{eq:hexp}
\end{equation}
The effective strain $E_{\text{eff}}$ is a shear-dominated strain measure indicative of slip, derived from the in-plane components of $\mathbf{H}^{exp}$ \cite{Paupler1988G.0070168938}:
\begin{equation}
E_{\text{eff}} = \sqrt{\frac{1}{2}\left(H_{xx}^{exp} - H_{yy}^{exp}\right)^2 + \left(H_{xy}^{exp}\right)^2}.
\label{eq:Eeff}
\end{equation}

The slip system's contribution to the displacement gradient tensor, for a single slip system $\alpha$, is expressed as the product of the slip magnitude $\gamma^\alpha$ and the Schmid tensor $\mathbf{P}^\alpha$:
\begin{equation}
\mathbf{H}^\alpha = \gamma^\alpha \mathbf{P}^\alpha = \gamma^\alpha \vec{s}^\alpha \otimes \vec{n}^\alpha.
\label{eq:Ha}
\end{equation}
Here, $\vec{s}^\alpha$ and $\vec{n}^\alpha$ represent the normalized slip direction and slip plane normal of slip system $\alpha$, respectively. When considering multiple active slip systems, the local theoretical displacement gradient tensor $\mathbf{H}^{theo}$ is the sum of the contributions of each system:
\begin{equation}
\mathbf{H}^{theo} = \sum_{\alpha=1}^N \mathbf{H}^\alpha.
\label{eq:Htheor}
\end{equation}

The original SSLIP method \cite{Vermeij2022AutomatedData} is predicated on the assumption that the measured in-plane displacement gradient tensor components can be described by a linear combination of Schmid tensors of slip systems, each with a certain slip magnitude (which can be 0, if the system is inactive). The optimization problem, defined at every datapoint, aims to minimize the sum of the absolute values of the slip activities, while constraining the tensor residual norm $R_{2D}^{L_{2}}$ below a predefined threshold $H_{thresh}$ (based on DIC noise level):
\begin{mini!}|1|
{\gamma^{\alpha}=\gamma^{1},\ldots,\gamma^{N}}{\sum_{\alpha=1}^{N}{|\gamma^{\alpha}|}\label{eq:objfun}}
{\label{eq:optim}}{}
\addConstraint{||\mathbf{H}^{exp}-\mathbf{H}^{theor}||_{2D}}{<H_{thresh}.\label{eq:con1}}{}
\end{mini!}

The minimization of the sum of absolute values of slip activities is included in order to handle more than four slip systems, since there are only four measured (known) components of the in-plane 2D displacement gradient tensor. In the original SSLIP paper \cite{Vermeij2022AutomatedData}, we showed that this approach works well for a less complicated HCP virtual experiment and also for FCC and BCC real experiments. However, as shown in \autoref{fig:HCPvirtualexperiment}, application to the challenging HCP virtual experiment with 4 overlapping slip systems from different slip families results in erroneous results.

\subsection{\psslip part I: Automated Preselection Using the Radon Transform}
\label{sc:automatedRa}

\begin{figure}[ht!]
    \centering
 \includegraphics[width=0.75\textwidth]{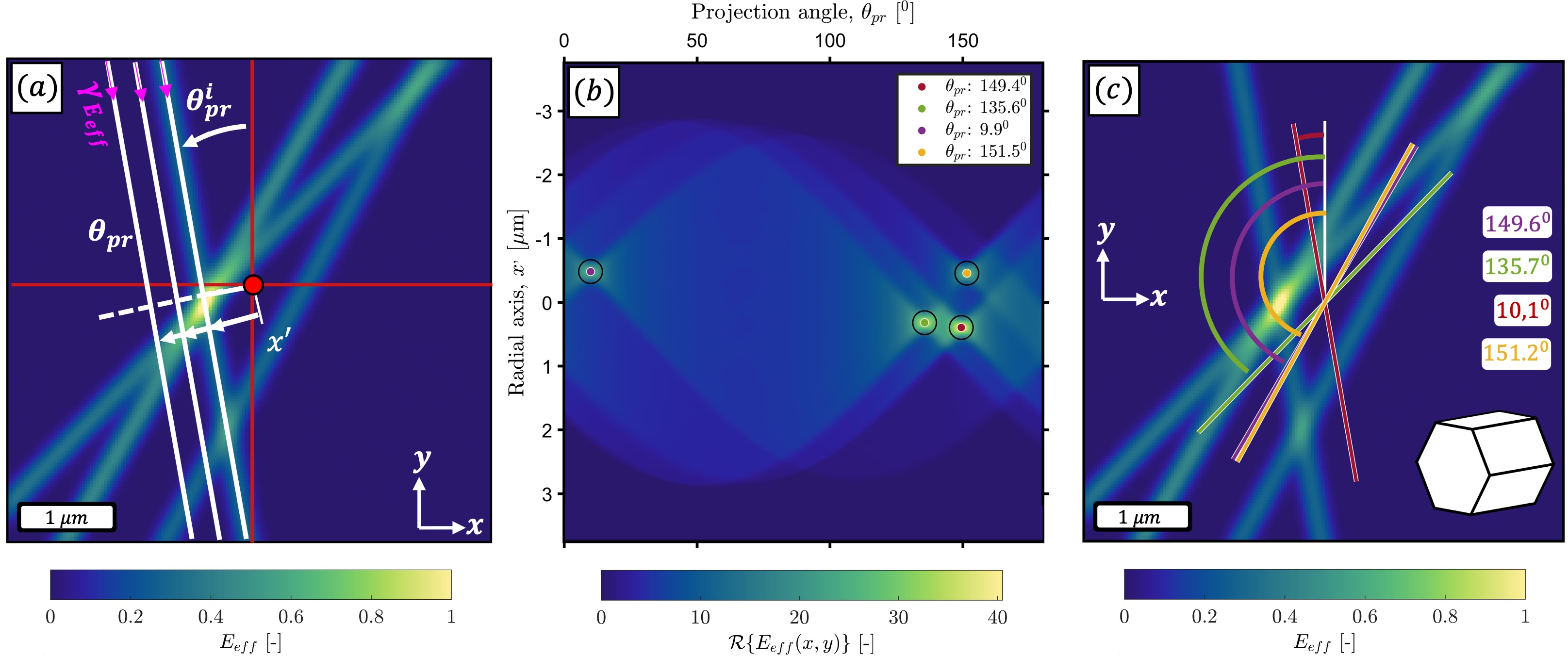}
    \caption{Application of the Radon transform to the Virtual HCP Experiment. (a) Effective strain field, $E_{\text{eff}}$, with superimposed lines indicating the projection angles, $\theta_{\text{pr}}$, used in the Radon transform. The intersection point $x_i$ marks the center of the local area considered for analysis. The red dot at the center indicates the rotation axis for the Radon transform. (b) The Radon transform output, $R\{E_{\text{eff}}(x, y)\}$, with peaks corresponding to prominent slip band angles in the strain field. (c) Identified slip trace orientations superimposed onto the strain field, with color-coded lines representing different slip system, detected by the Radon transform, with corresponding theoretical angles.}
    \label{fig:Ratransformbasicsvisualisation}
\end{figure}

To increase the robustness, the \psslip methodology first incorporates a new step for automated preselection of slip systems through the Radon transform, as proposed by Hu \textit{et al.} \cite{Hu2023AutomatedAlloy}. An important difference with the method of Hu \textit{et al.} is that we use the Radon transform only for deselection of slip systems that clearly cannot be active, while the actual identification is performed in the second part of the \psslip method, as discussed below. The Radon transform integrates a 2D function, in this case the measured effective strain field $E_{eff}$, along a single line (limited to $x' = x \cos \theta_{pr} - y \sin \theta_{pr}$ by usage of the Dirac delta function $\delta$) and projects that value into Radon space ($\mathcal{R}\{E_{eff}(x, y)\}$), which is spanned by the variable $\theta_{pr}$ and  $x'$  \cite{helgason1999radon, Hu2023AutomatedAlloy}:
\begin{equation}
\mathcal{R}\left\{E_{eff}(x, y)\right\} = \iint_D E_{eff}(x, y) \delta\left(x' - x \cos \theta_{pr} - y \sin \theta_{pr}\right) dx dy.
\label{eq:Radontransformbasics}
\end{equation}
Here, $\theta_{pr}$ is the positive angle of the line with respect to the original y-axis, while $x'$ denotes the smallest distance between the line and the origin of the cartesian coordinate system. Line integrals are performed along all possible lines across the given data set, which yields a sinogram that highlights dominant slip bands as high intensity peaks, as shown in \autoref{fig:Ratransformbasicsvisualisation}(a,b). The measured slip band angles are then matched to theoretical slip trace orientations, using a threshold of $\pm$10~\degree.
In \autoref{fig:Ratransformbasicsvisualisation}(c), the identified slip traces are shown with their corresponding angles, which agree perfectly with the measured ones in \autoref{fig:Ratransformbasicsvisualisation}(c).

\begin{figure}[ht!]
    \centering
 \includegraphics[width=1.0\textwidth]{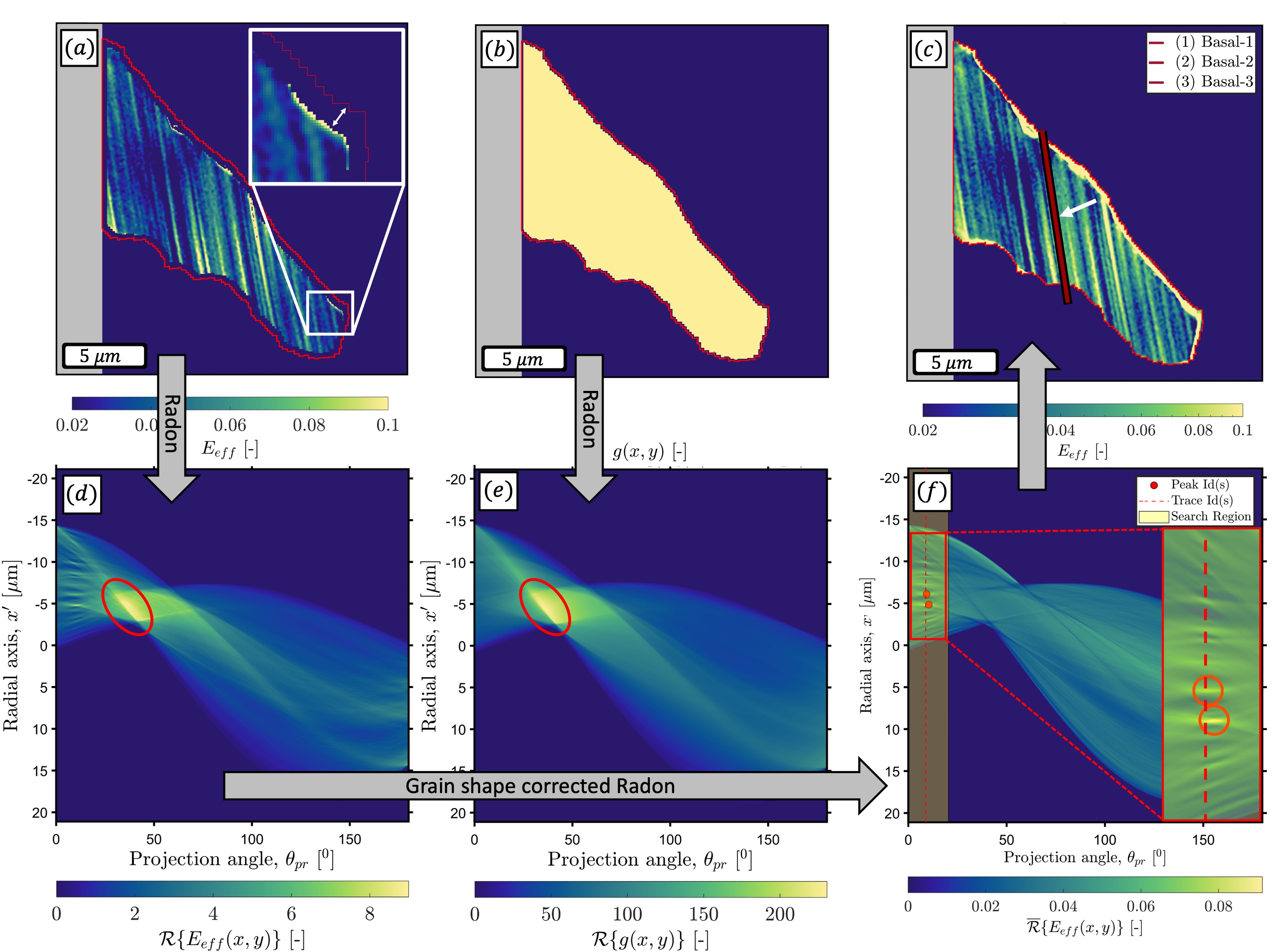}
    \caption{Experimental grain analysis Using Normalized Radon transform. (a) Effective strain field within a single grain, $E_{\text{eff}}$, with a magnified inset detailing the grain boundary, around which data is removed. (b) Binary mask derived from the strain field to define the grain boundary for normalization. (c) Superimposed identified Basal slip trace on the strain field, conforming to the identified Radon transform peaks. (d) Raw Radon transform, $R\{E_{\text{eff}}(x, y)\}$, highlighting the maximum values with a red ellipse. (e) Grain shape Radon transform, $R\{g(x, y)\}$, which accounts for the curvature of the grain boundary, showing a similar profile to that of (d). (f) Corrected Radon output with peak indicators (Magnified in the inset) and trace lines corresponding to the detected slip bands within the search region, illustrating the precision of slip band orientation identification post grain shape normalization.}
    \label{fig:Ratransform}
\end{figure}

While this works well on the virtual experiment, we try this now on an experimental strain field to assess its performance. In order to avoid the influence of grain boundary deformation, we first remove a band of 5 pixels along the grain boundary. \autoref{fig:Ratransform}(a) shows the strain field of a grain from the dataset of the case study in \autoref{sec:hcp_case_study}. Its Radon transform, shown in \autoref{fig:Ratransform}(d), shows a clear maximum as illustrated with the red ellips, which, as will be shown, does not correspond to the actual slip band orientations. Therefore, we address the influence of the grain shape by introducing a correction step, similar to the approach by Hu \textit{et al.} \cite{Hu2023AutomatedAlloy}. A binary mask of the grain shape (\autoref{fig:Ratransform}(c)),
\begin{equation}
g(x, y) = 
\begin{cases}
1, & \text{for }(x, y) \text{ inside the grain} \\
0, & \text{for }(x, y) \text{ outside the grain}
\end{cases},
\label{eq:shapefunction}
\end{equation}
is employed to normalize the Radon transform outputs. The Radon transform of this grain shape function $g(x,y)$ ($\mathcal{R}\{g(x, y)\}$) is shown in \autoref{fig:Ratransform}(e). The same Radon maximum values are seen as in \autoref{fig:Ratransform}(d), which confirms that the slip bands are not identified. Next, normalization of the peak heights in the original Radon space is performed:
\begin{equation}
\overline{\mathcal{R}}\left\{E_{eff}(x, y)\right\}=\frac{\mathcal{R}\left\{E_{eff}(x, y)\right\}}{\mathcal{R}\{g(x, y)\}}.
\end{equation}
This ensures that the length of integration reflects the actual slip trace length within the grain, as shown in \autoref{fig:Ratransform}(f), wherein the Radon peaks actually correspond to the slip bands. 

Slip band orientations are identified by finding peaks in the normalized Radon sinogram (\autoref{fig:Ratransform}(f)). We employ the \texttt{peaks2} function iteratively in Matlab \cite{peaks2}. \texttt{peaks2} operates by comparing each data point to its eight immediate neighbors in a vectorized manner, ensuring efficient execution. A point is considered a peak if it is strictly greater than all its surrounding neighbors. Relevant parameters are summarized in \autoref{tab:radonPeaks}. Starting with a minimum peak height of 1 (maximum theoretically possible), the minimum peak height is lowered in steps of $IncPeakHeight$ until the desired number of peaks ($NmaxPeaks$) are identified, or until the overall minimum peak height ($MinPeakHeight$) is reached. The peaks need to be spaced at least $MinPeakDistance$ pixels. If multiple peaks are identified within that range, then the one with the highest intensity is chosen as the only peak. The slip band angles corresponding to the identified peaks are then compared to theoretical slip trace angles (from EBSD) and must fall within a threshold of $OrientThresh$ to be considered a potential slip system. The result is shown in \autoref{fig:Ratransform}(f), in which two peaks are identified. Note that the next highest peaks have the same angle, meaning that the net result would be the same, which we found is often the case. Moreover, missing several minor peaks of one slip system is not an issue, since the purpose is only to identify the presence of active slip systems using the Radon, not to identify their location or amount of active bands. For both reasons, the precise peak finding parameters are not very important, as far as we can tell. Finally, \autoref{fig:Ratransform}(c) shows the strain map again with the identified (Basal) slip trace.

\begin{table}[H]
\centering
\caption{Parameters for peak identification in Radon transform}
\label{tab:radonPeaks}
\begin{tabular}{lccr}
\hline
\textbf{Parameter} & \textbf{Value} & \textbf{Unit} \\
\hline
$MinPeakHeight$  & 0.8 & [-] \\
$NmaxPeaks$  & 5 & [-] \\
$MinPeakDistance$ &   50 & [pix] \\
$IncPeakHeight$ &   0.025 & [-] \\
$OrientThresh$ &   10 & [\degree] \\
\hline
\end{tabular}
\end{table}

\begin{figure}[ht!]
    \centering
 \includegraphics[width=0.9\textwidth]{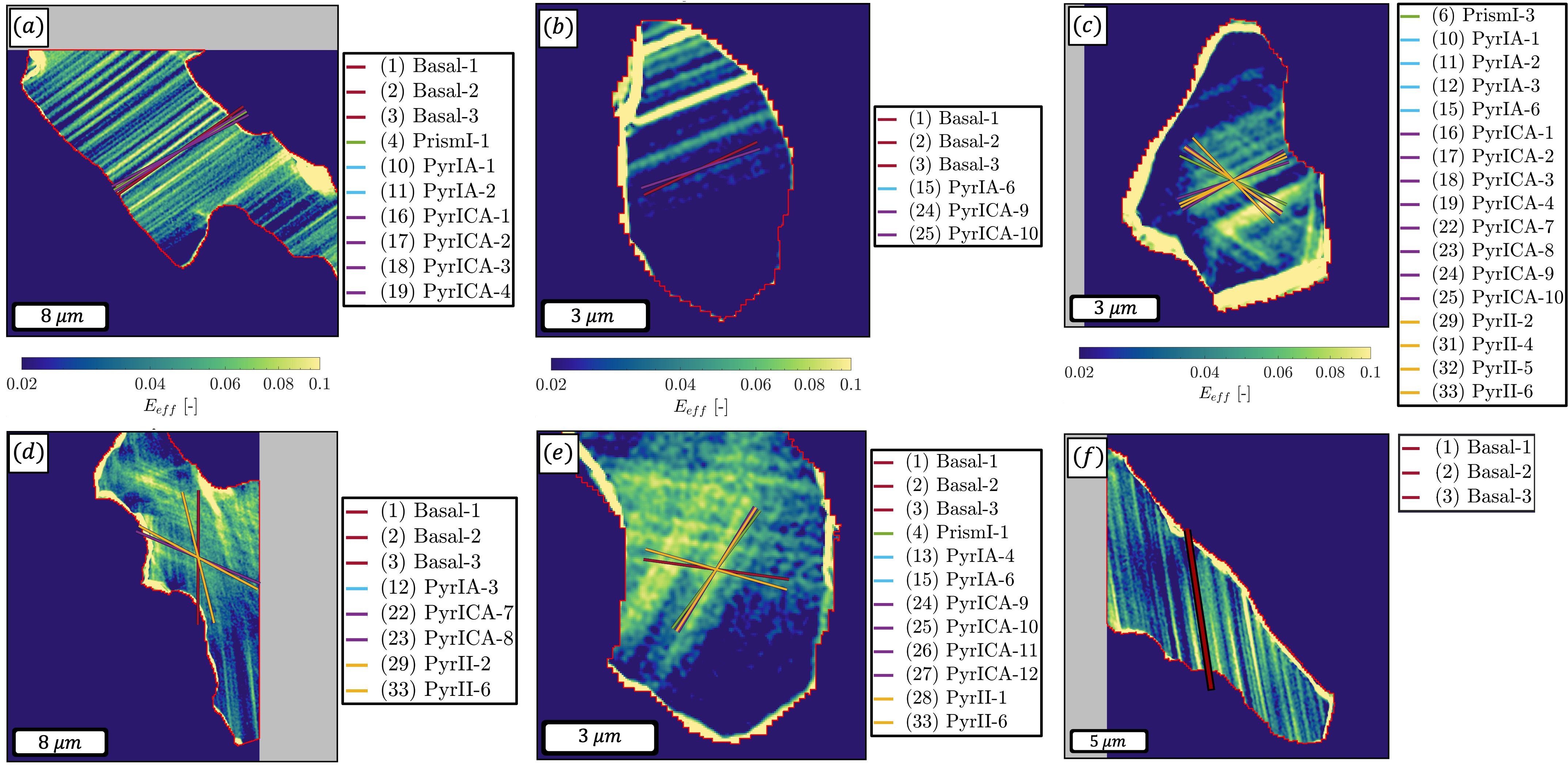}
    \caption{Demonstration of Radon transform preselection on all grains considered in this work. For each grain, (a-f), the effective strain field $E_{\text{eff}}$ is plotted. The slip traces of the preselected slip systems are drawn as overlay, as described in the legends. Note that the PyrIA slip traces are not visible since they overlap with the PyrICA slip traces.}
    \label{fig:RatransformAll}
\end{figure}

As a preliminary demonstration, we show in \autoref{fig:RatransformAll} the Radon transform preselected slip systems for all experimental cases considered in this paper. The simple example grain shown in \autoref{fig:Ratransform} resulted in preselection of only one single slip family, i.e. Basal slip, see \autoref{fig:RatransformAll}(f). In such cases, it could be argued that further identification with \psslip is not required, since a Schmid Factor (SF) based method of, e.g., Hu et al. \cite{Hu2023AutomatedAlloy}, would perform well here in the case that the slip system with the highest SF within a slip family is always activated. However, the SF is known to be unreliable for polycrystalline microstructures \cite{chen_daly_2016}. Additionally, for smaller grains with fewer dislocation sources, the activation of slip systems is more stochastic, making the SF less useful. Moreover, \autoref{fig:RatransformAll}(a-e) shows that all other grains considered in this work show at least 3 preselected slip families (at least 6 slip systems for each grain). Therefore, it is unreasonable in these cases to rely just on the SF for further selection, especially when the critical resolved shear stresses can be unknown and different for the different slip families, as is often the case. Additionally, while \autoref{fig:RatransformAll}(a,b) show clear slip bands that are visually conforming well to the preselected slip traces, the initial Radon-based preselections of the grains in \autoref{fig:RatransformAll}(c-e) are less conclusive. In summary, the Radon transform is a valuable first step, but the SSLIP analysis is required to achieve positive identification of the precise slip systems and their activity fields.

\subsection{\psslip part II: SSLIP Analysis with Preselected Slip System Combinations}
\label{sc:methodSSLIP}

With the Radon transform providing a trace-based preselected set of slip systems, we can now apply SSLIP in a targeted manner. For improved robustness and to avoid erroneous solutions from in-plane linearly dependent slip systems, we follow a refined approach that involves systematically examining every possible combination of two slip systems from the Radon transform preselected pool (of $N_{radon}$ slip systems), at each datapoint. Essentially, this means conducting the SSLIP analysis $N_{duo}$ times for every datapoint, where $N_{duo}$ represents the total number of unique two-system combinations (i.e. duos) from the preselected set:
\vspace{-0.5em}
\begin{equation}
    N_{duo} = \frac{N_{radon}!}{2!*(N_{radon}-2)!}.
\label{eq:N_duo}
\end{equation}
For the virtual experiment introduced earlier, preselection excludes 16 slip systems, only leaving 3 Basal systems, 1 Prismatic system, 2 PyrICA systems and 2 PyrII systems to be included ($N_{radon}$ = 8). Thus, according to \autoref{eq:N_duo}, this results in $N_{duo}=28$ combinations to be analyzed by SSLIP. Below, we will detail the criteria and processes for selecting the optimal combination for each datapoint, ensuring a precise and comprehensive identification of active slip systems.

\subsubsection{SSLIP Analysis with Preselected Slip System Combinations}
\label{sc:sslip_combinations}

For each pair of slip systems, we now apply the SSLIP methodology, utilizing the optimization equation (\autoref{eq:optim}) and also correcting for rotations, as explained below in \autoref{sec:rotCor}. Subsequently, the selection of the optimal combination for each datapoint is governed by the residual fraction, $Rf_{2D}^{L_{2}}$, which must fall below a predefined threshold to ensure accurate identification. This residual fraction is calculated as
\begin{equation}
Rf_{2D}^{L_{2}} = \frac{\left\|\mathbf{H}^{exp}-\mathbf{H}^{theo}\right\|_{2D}}{E_{eff}},
\label{eq:residualfraction}
\end{equation}
where $\mathbf{H}^{exp}$ represents the experimental displacement gradient tensor, and $\mathbf{H}^{theo}$ denotes the theoretical displacement gradient tensor predicted by a combination of slip systems and rotation (see \autoref{sec:rotCor}), and $E_{eff}$ is the effective strain, serving as a normalization factor.

The constraints imposed on potential slip system combinations are outlined in \autoref{tab:SSLIP_Parameters}.
\begin{table}[H]
\centering
\caption{Parameters for SSLIP Analysis Selection Criteria}
\label{tab:SSLIP_Parameters}
\begin{tabular}{lccr}
\hline
\textbf{Parameter} & \textbf{Symbol} & \textbf{Value} & \textbf{Unit} \\
\hline
Max. Displacement Gradient Tensor Residual Fraction & $Rf_{2D,max}^{L_{2}}$ & 0.2 & [-] \\
Min. Equivalent Strain & $E^{min}_{eff}$ & 0.02 & [-] \\
Max. Allowable Rotation Correction & $\gamma^{max}_{rot}$ & 5 & [$^0$] \\
\hline
\end{tabular}
\end{table}
Each combination is thereby assessed based on its residual fraction (\autoref{eq:residualfraction}), rotation angle and minimum equivalent strain, according to the values specified in \autoref{tab:SSLIP_Parameters}. These values have been chosen as they were found to work robustly for the analysis of hundreds of Zn grains, of which a few challenging cases are discussed in \autoref{sec:hcp_case_study}. Among the potential combinations which satisfy these three criteria, the one with the lowest sum of slip magnitudes, thereby indicating the path of least resistance and conforming to the principle of minimum energy, is selected as the definitive combination of slip system activities for that particular datapoint.

\subsection{\psslip part III: Rotation Correction in SSLIP}
\label{sec:rotCor}

Since SSLIP will be applied with only two systems at a time, the optimization problem will have fewer degrees of freedom, which allows us to simultaneously identify the local elastic rotations. The inclusion of rotations was not needed with SSLIP for the less challenging cases of FCC and BCC, but is found to be necessary for HCP because of its large number of slip systems with linearly dependent in-plane deformation gradient tensors. In practice, these rotations can be caused by (full) grain rotation (e.g. due to grain boundary sliding), by compatibility effects near a grain boundary or by other microstructural features \cite{vermeij2022nanomechanical}, or simply by misalignment of the sample between (quasi \textit{in-situ}) measurement steps. Therefore, we add (local) rotation correction to the methodology, which will not only improve the slip system identification but will also give an extra result, i.e. the rotation field, which may be useful for further analysis.

The correction involves incorporating an additional rotation component into SSLIP, represented by an approximated rotation tensor. For small rotations, which are common in our observations, the in-plane rotation tensor can be approximated as follows:
\begin{equation}
\mathbf{H}^{rot} =
\begin{bmatrix}
\cos(\theta) & -\sin(\theta) \\
\sin(\theta) & \cos(\theta) \\
\end{bmatrix} - \mathbf{I} \approx \theta 
\begin{bmatrix}
0 & -1 \\
1 & 0 \\
\end{bmatrix}.
\label{eq:rotationcorrection}
\end{equation}
Here, $\theta$ represents the rotation angle and, under the assumption that $\theta$ is small, $\sin(\theta) \approx \theta$ and $\cos(\theta) \approx 1$, allowing us to capture the in-plane rotation with this approximated matrix, in displacement gradient tensor form. It is then incorporated into SSLIP as an additional "slip system", for each combination that will be assessed. This results in the rotation value $\theta$ at every datapoint, along with the slip system magnitudes. In cases where the rotations would be so large that the simplified rotation tensor is a poor estimate, the full version of $\mathbf{H}^{rot}$ in \autoref{eq:rotationcorrection} can be used, at the expense that $\mathbf{H}^{rot}$ needs to be updated using the current estimate of the rotation angle, at each location and at each step in the iterative optimization routine. After successful identification, the 2D kinematics at each data point is described as
\begin{equation}
\mathbf{H}^{theor} = \gamma^i \mathbf{P}^i_{2D} + \gamma^j \mathbf{P}^j_{2D} + \mathbf{H}^{rot},
\end{equation}
in which $i$ and $j$ are the two identified slip systems.

\subsection{Validation on Virtual Experiment}
\begin{figure}[ht!]
    \centering
 \includegraphics[width=0.95\textwidth]{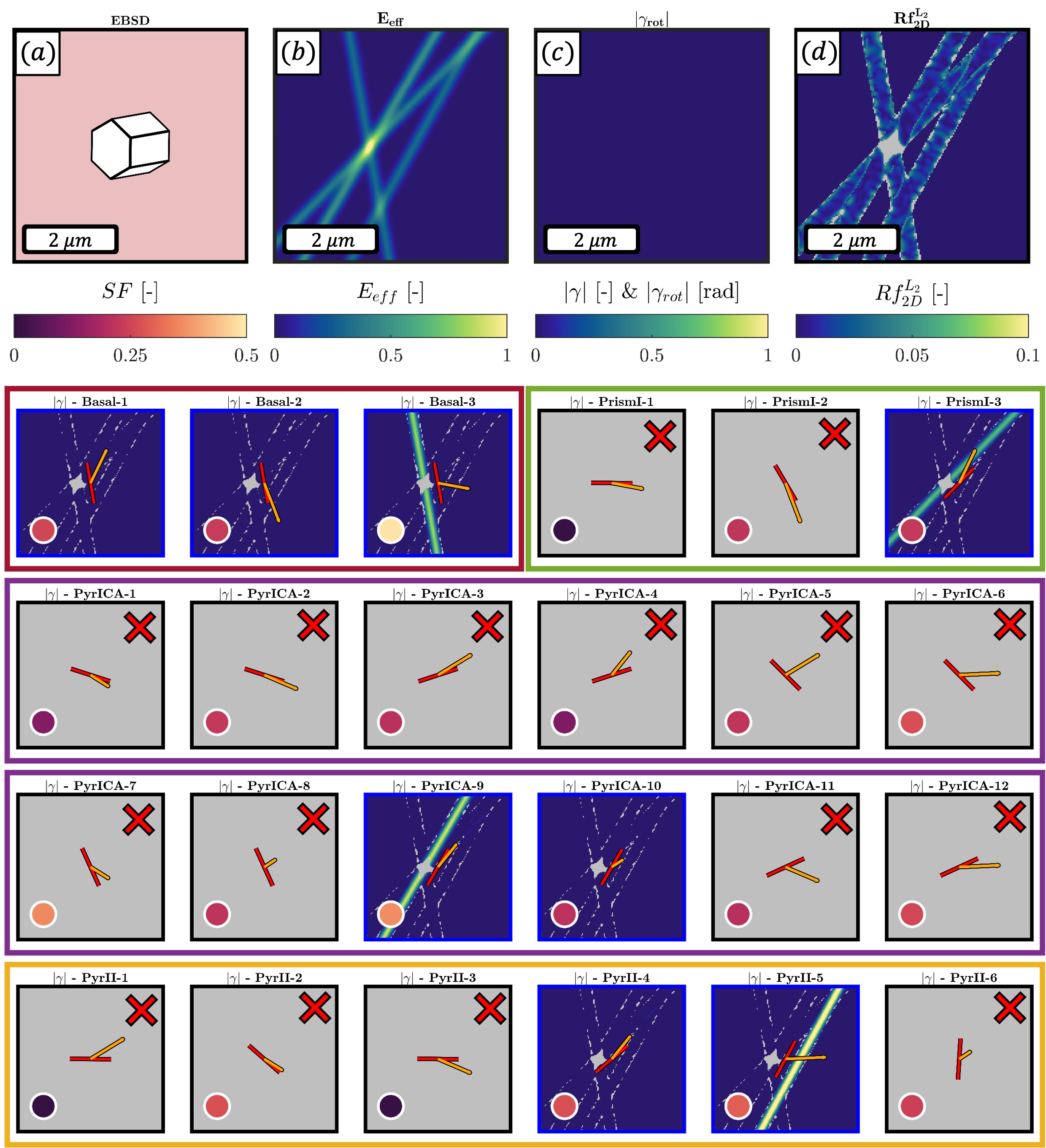}
    \caption{Demonstration of the \psslip results on the challenging virtual experiment. (a) The EBSD map and crystal shape illustrate the crystal orientation of this virtual experiment. (b) The effective strain field, $E_{eff}$, clearly showing four slip bands. (c) The rotation field, $\gamma_{rot}$, resulting from rotation-correction by \psslip. (d) The residual strain fraction field,  $Rf_{2D}^{L_{2}}$, resulting from \textbf{+}SSLIP, showing the fit between the "experimental" and theoretical displacement gradient tensor. Below that, the slip system activity map of each slip system of each slip family (Basal, PrismI, PyrICA and PyrII) is shown. Maps with red crosses indicate that systems were not selected by Radon transform preselection. Each map also features the slip trace (red line) and direction (orange arrow), with the Schmid factor ($SF$) depicted as colored circles (colorbar below (a)). The SF is not used in the \psslip method, but plotted here such that structure of the figure corresponds to the experimental case studies}
    \label{fig:RRSSLIPProcess}
\end{figure}

Now, the full \psslip methodology is applied to the challenging HCP virtual experiment, for which the results are shown in \autoref{fig:RRSSLIPProcess}. This figure is structured identically to those in the experimental case studies below. \autoref{fig:RRSSLIPProcess}(a) shows the crystallographic orientation of the virtual experiment and \autoref{fig:RRSSLIPProcess}(b) shows the effective strain field. Application of the Radon transform results in eight preselected slip systems, see \autoref{fig:Ratransformbasicsvisualisation}. Subsequently, application of Rotation-corrected SSLIP on all pairs of the slip systems ($N_{duo}=28$) and point-wise selection of the pair with the lowest residual, results in the slip activity fields as shown in \autoref{fig:RRSSLIPProcess}. The correct slip systems were Basal-3, PrismI-3, PyrICA-9 and PyrII-5, all of which show a single slip band in their respective slip activity field, oriented as expected along their slip trace. The other four preselected but non-active slip systems show no activity over the full area, as expected. The error of the SSLIP analysis is represented by the residual norm fraction in \autoref{fig:RRSSLIPProcess}(d), $Rf_{2D,max}^{L_{2}}$ (\autoref{eq:residualfraction}), for which low values are observed. At the edges of the slip bands, the errors are higher because the error is normalized with the effective strain, which is low in these areas, leading to a lower signal to noise ratio. No rotation was applied in the virtual experiment, which is reproduced by the \psslip algorithm as shown in \autoref{fig:RRSSLIPProcess}(c). The intersection of three slip bands in the center of the domain cannot be identified (grey area), since only two slip systems can be active per datapoint. Importantly, these intersections are also not falsely attributed to incorrect slip systems. In practice, in the unlikely case when more than two systems overlap, the slip activity at such multi-slip intersections can easily be inferred from the neighbouring slip activity. Overall, the identification shown here is robust and accurate and shows promise for identification of complex HCP activities. 

\subsection{Conclusive vs. Inconclusive Identification}
\label{subSec:inconcl}

The \psslip method is demonstrated to accurately identify slip system activities on the virtual experiment, however, when many slip systems are considered, the kinematics of certain combinations of systems may not be distinguishable in 2D (i.e. in-plane). In theory, even though each slip system has a unique 3D displacement gradient, the 2D part of the displacement gradient tensor could be exactly, or almost, the same for two (or more) slip systems, making those slip systems fundamentally nondiscriminatory in the SSLIP method. In practice, however, the minimization of the sum of slip activities, as used in the original SSLIP method \cite{Vermeij2022AutomatedData}, was found to be robust against slip systems with almost equal 2D displacement gradient tensors, therefore, this issue was not considered in detail before. For HCP, however, there are many potential slip systems from multiple different families, which typically cannot be excluded a-priori due to unknown Critical Resolved Shear Stress (CRSS) values. As a result, the occurrence of multiple highly similar 2D displacement gradient tensors was found to be problematic for certain grain orientations, a few examples of which are treated in \autoref{subSec:conclusivevsinconclusive}. In such cases, a group of two or more slip systems can be defined with similar enough 2D displacement gradient tensors such that the measured slip activity can only be assigned to the group of slip systems. When slip activity is assigned to such a 'nondiscriminatory slip system group', we call the identification 'inconclusive', as counterpart to the regular 'conclusive' identification where the slip activity is uniquely assigned to one or more slip systems. 

Therefore, in the \psslip method, after the Radon transform preselection, all possible nondiscriminatory slip system groups are identified from the available preselected slip systems, before executing the \psslip identification. The likelihood of misidentification due to the existence of slip systems that are similar in their 2D kinematics, depends on the absolute measurement accuracy of the slip trace angle and the projected slip direction angle. In turn, these depend on the combined accuracy of EBSD, SEM-DIC (including SEM scanning artefacts), and their mutual alignment. In this work, it was found that these two angles can have an error up to $\sim$5 \degree. Therefore, we deﬁne an angle threshold of 5 \degree, i.e., if two preselected slip systems have both a slip trace angle and projected slip direction angle within 5 \degree of each other, they are considered nondiscriminatory. Through this procedure, the preselected slip systems are, when necessary, included in nondiscriminatory slip system groups in the \psslip identification. These groups can consist of more than two slip systems. In the (pixel-wise) \psslip output, the slip activity assigned to a nondiscriminatory group is attributed equally to each slip system within the group. The way this works in practice will be shown in the section below (note that the virtual test case in \autoref{fig:RRSSLIPProcess} is a regular example of a 'conclusive' identification).

\section{Experimental Case Study: \psslip applied on HCP Zinc Coating}
\label{sec:hcp_case_study}

In this section, we demonstrate the strength of the \psslip method on experimental SEM-DIC results of an HCP Zinc anti-corrosion coating, which will show activities from multiple slip families: Basal, PyrI and PyrII. The focus will be on challenging case studies that include multiple 'nondiscriminatory slip system groups'. Additionally, we will showcase robust identification on grains that exhibit cross slip and diffuse slip, for which pure trace-based analysis is hardly possible \cite{Vermeij2022AutomatedData}.

In \autoref{subsec:material_char_deformation}, we briefly detail the experimental methodology for acquiring microstructure-aligned SEM-DIC strain fields. \autoref{subsec:singleSlip} demonstrates \psslip on a simple case of single slip and \autoref{subSec:conclusivevsinconclusive} shows how an inconclusive identification can be interpreted. More complex cases of (diffuse) multi-slip and cross slip are treated in Sections \ref{subSec:multiSlip} and \ref{subSec:complexSlip}.

\subsection{Material Characterization and SEM-DIC Testing}
\label{subsec:material_char_deformation}

The material used in this case study is a hot-dip galvanized skin-passed Zn coated steel sample with material code DX54-Galvanized Iron (GI). Samples are cut using wire-EDM (gauge width 4 mm, thickness 0.7 mm). Before testing, the sample surface is carefully prepared in a single step of polishing using Oxide Polishing Suspension (OPS)-NonDry and ethanol (1:1) suspension for 10 minutes. This short polishing provides the optimal compromise between good surface quality and low material removal, since the Zn coating is only $\sim10 \mu m$.

\begin{figure}[htb!]
    \centering
    \includegraphics[width=\textwidth]{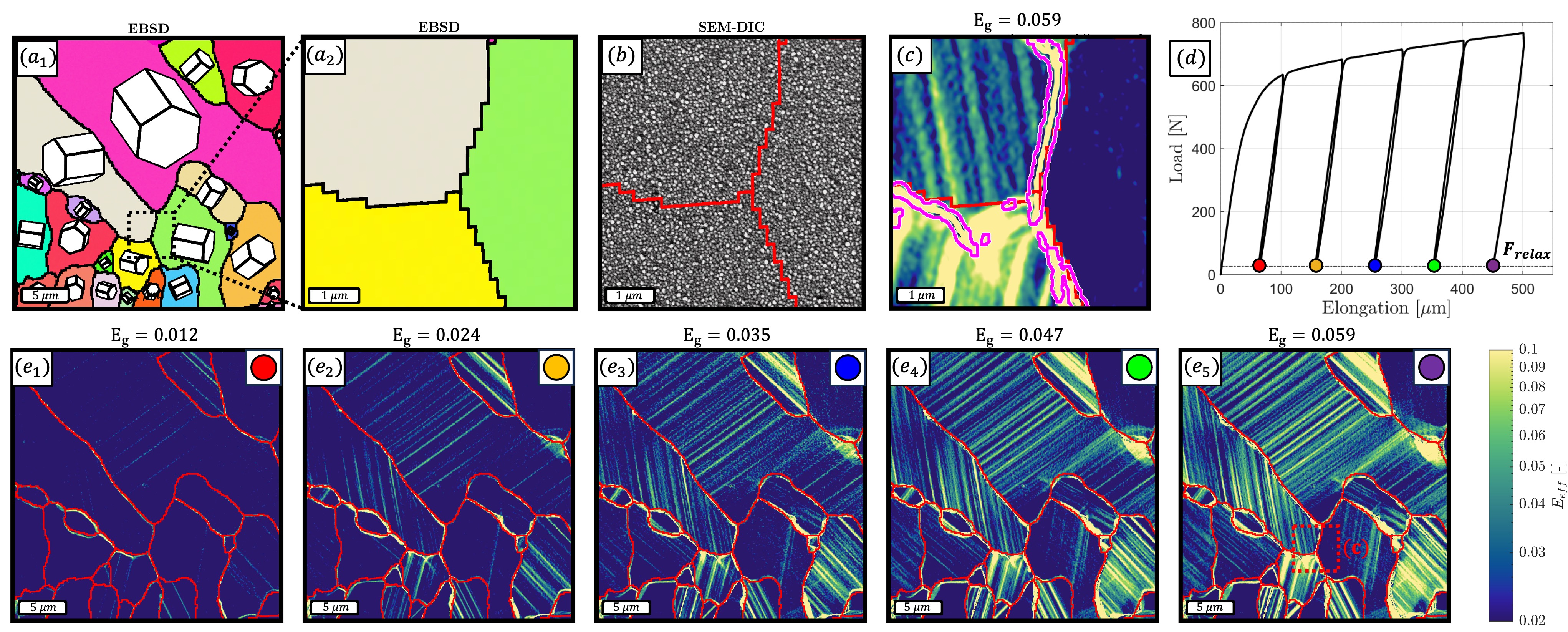}
    \caption{(a\textsubscript{1}) EBSD (out-of-plane) IPF map  illustrating the microstructure of the Zn coating prior to deformation, with hexagonal shapes illustrating the crystal orientations. (a\textsubscript{2}) Zoom-in of IPF map around a triple junction. (b) Scanning Electron Microscope (SEM) in-lens Secondary Electron (SE) scan displaying the Indium-Tin (InSn) speckle pattern used for Digital Image Correlation (DIC). The aligned grain boundaries are shown as overlay in red. (c) Effective strain field captured (inset of (e\textsubscript{5}) via DIC at a global strain of 0.059, showing crystallographic slip and grain boundary deformation. Areas surrounded by pink lines denote interpolated DIC data, predominantly at grain boundary locations. (d) Load-elongation curve from the \textit{in-situ} tensile test with color-coded points corresponding to the strain levels in subfigures (e\textsubscript{1})-(e\textsubscript{5}). (e\us{1})-(e\us{5}) Effective strain fields at progressive tensile test increments, highlighting the evolution of strain localization and the alignment with EBSD data.}
    \label{fig:experiment_overview}
\end{figure}

Next, several $\sim$50 $\mu$m sized Regions of Interest (ROIs) are chosen for characterization and \textit{in-situ} SEM-DIC tensile testing, as shown in \autoref{fig:experiment_overview} for one ROI. EBSD is performed with an EDAX Digiview 2 camera, with offline spherical indexing using EMSphInx for improved quality \cite{LENTHE2019112841}, resulting in an Inverse Pole Figure (IPF) map as visualized in \autoref{fig:experiment_overview}(a). Subsequently, we apply a fine and dense InSn SEM-DIC speckle pattern applied according to the parameters of pattern \textbf{b} in Table 1 of Hoefnagels \textit{et al.} \cite{HoefnagelsPattern}, resulting in approximately 90-100 nm sized speckles, see \autoref{fig:experiment_overview}(b). The specimen is then mounted in a Kammrath\&Weiss micro-tensile stage, which is installed inside a Tescan Mira 3 SEM for interrupted \textit{in-situ} testing, at deformation steps as shown in \autoref{fig:experiment_overview}(d). For SEM-DIC imaging, we use in-lens SE imaging (5 kV beam voltage at 7 mm working distance) \cite{Vermeij2021, vermeij2022nanomechanical, vermeij2023damage}.
At each deformation step, a horizontal and vertical scan are captured that are combined into a single image, thereby correcting for SEM scanning artefacts, using \textit{ScanCorr} \cite{vermeij2022nanomechanical}. Subsequently, DIC is performed on these corrected images using MatchID DIC software, employing parameters as given in \autoref{tab:dic_parameters}, following procedures proposed by Vermeij \textit{et al.} \cite{Vermeij2021, vermeij2022nanomechanical}. Subsequently, a nanomechanical alignment framework \cite{vermeij2022nanomechanical} is applied to allow direct spatial correlation between EBSD and DIC data, as observed in the strain maps of all deformation steps in \autoref{fig:experiment_overview}(e), with the aligned grain boundaries overlaid. 

At first glance, the material reveals a range of deformation mechanisms. Crystallographic slip can be observed within grains, both in the form of sharp and more diffuse slip bands. Additionally, localizations occur on top of grain boundaries, hinting at the occurrence of grain boundary sliding and/or migration. \autoref{fig:experiment_overview}(c) shows this more clearly with an inset of the strain field around a triple junction, for the final deformation step. Discrete slip, diffuse intragranular deformation, and grain boundary activity can be observed. Note that the grain boundary deformation is very strong and resulted in a local degradation of the DIC pattern, such that interpolation of the DIC data was required (in the areas outlined with a thin pink line), as described in \cite{vermeij2023damage}. Grain boundary deformation is outside the scope of this paper and will be subjected to identification by an alternative variation of the SSLIP method focused on grain boundary sliding \cite{Mornout2024LIBS}.

\begin{table}[htb!]
\centering
\caption{DIC System and Correlation Parameters}
\label{tab:dic_parameters}
\begin{tabular}{lcc}
\hline
\textbf{Parameter} & \textbf{Value} & \textbf{Unit} \\
\hline
Capture Instrument & Tescan Mira 3 SEM & - \\
Field of View & 40 &  [$\mu$m] \\
Pixel Size & 13 & [nm] \\
DIC Software & MatchID & - \\
Correlation Algorithm & ZNSSD & - \\
Subset Size & 33 & [pix] \\
Step Size & 3 & [pix] \\
Matching Criterion & Zero-normalized sum of squares differences & - \\
Pre-filtering & Gaussian & - \\
Filter Size & 1 & [pix] \\
Correlation Threshold & $\geq 0.7$ & - \\
Subset Shape Function & Affine & - \\
\hline
\end{tabular}
\end{table}

With the microstructure-correlated deformation data available, we will now demonstrate the \psslip method on several HCP Zn grains. For this, we choose several individual grains that show slip activity, starting with simple Basal slip and moving on to multiple and differently oriented slip bands in \autoref{subSec:multiSlip}. Additionally, "nondiscriminatory" slip systems will be handled transparently when encountered. In \autoref{subSec:complexSlip}, we will focus on more complex cases of cross slip and diffuse slip, on which the \psslip method excels.

\subsection{\psslip applied on Zn: Single Slip}
\label{subsec:singleSlip}

\begin{figure}[htb!]
    \centering
    \includegraphics[width=0.8\textwidth]{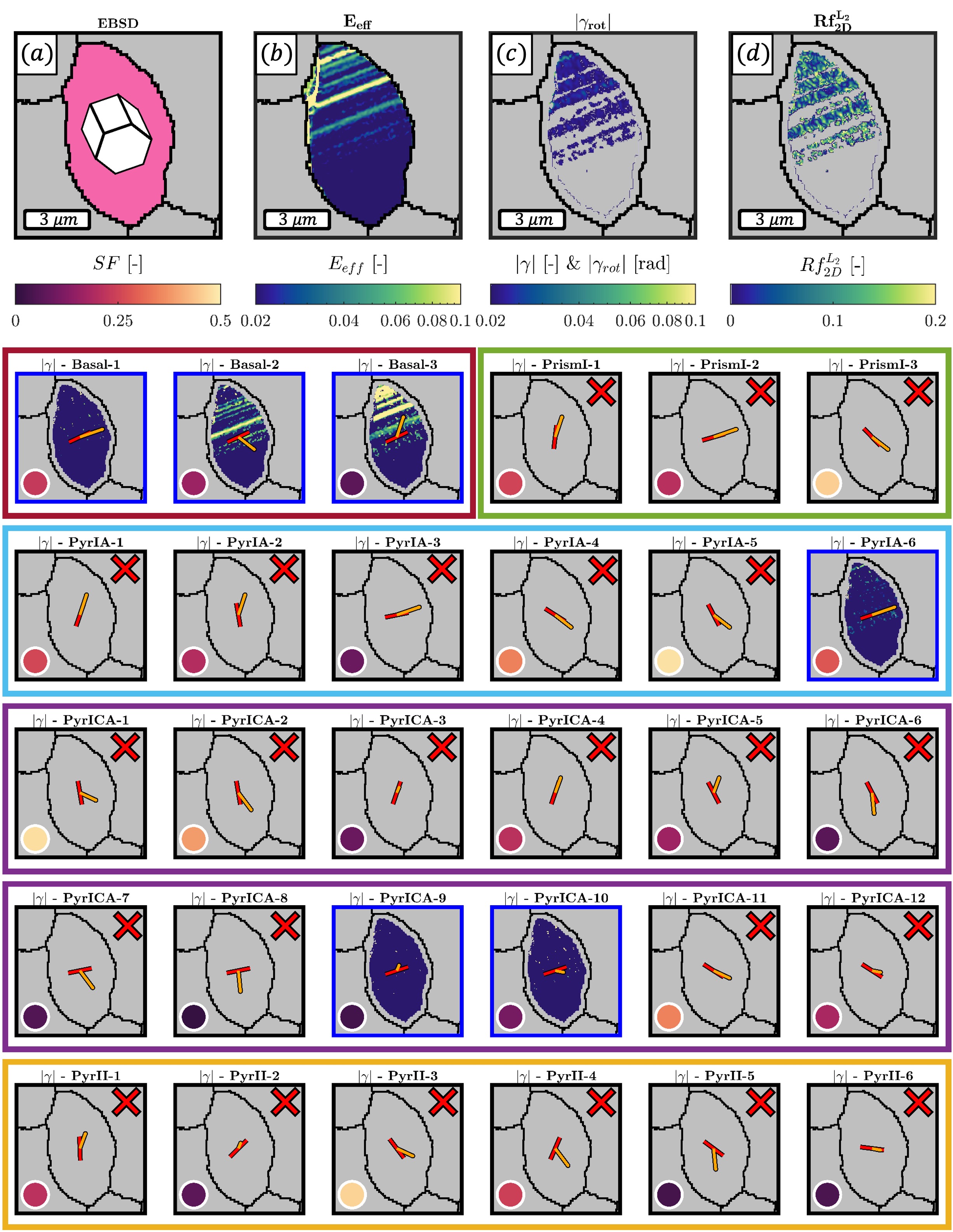}
    \caption{Results of slip system identification for a selected HCP Zn grain, showing single slip, using the \psslip method. (a) The EBSD map and rotated crystal shape illustrates the crystal orientation. (b) The effective strain field, $E_{eff}$, showing the deformation pattern. (c) The rotation field, $\gamma_{rot}$, resulting from rotation-correction by \psslip. (d) The residual strain fraction field,  $Rf_{2D}^{L_{2}}$, resulting from \psslip, showing the fit between the experimental and theoretical displacement gradient tensor. Below that, the slip activity map of each slip system of each slip family (Basal, PrismI, PyrIA, PyrICA and PyrII) is shown. Maps with red crosses indicate that systems were not selected by Radon transform preselection. Each map also features the slip trace (red line) and direction (orange arrow), with the SF depicted as colored circles (colorbar below (a)). Note that the SF is not used in the \psslip method.}
    \label{fig6}
\end{figure}

The first experimental grain that we investigate is shown in \autoref{fig6}. The crystallography and the deformation of the grain (at the final deformation increment) is given in \autoref{fig6}(a-b). At a first glance, slip occurs over only one specific slip plane, making it a relatively simple case to start the demonstration of the \psslip methodology. 

Application of \psslip results in identification of the slip activity fields as shown in \autoref{fig6}. As expected, the Radon transform preselection only picks up one specific slip band, which in this case matches with six individual slip systems: all 3 Basal systems, 1 PyrIA system and 2 PyrICA systems. Among these, Basal-2 and Basal-3 are identified as the active slip systems, which is unsurprising since Basal slip is known to have the lowest CRSS in Zn \cite{Parisot2004DeformationModes, bednarczyk2023determination}. The rotation field in \autoref{fig6}(c), as identified by \psslip, shows a negligible contribution for this grain. Additionally, the residual fraction in \autoref{fig6}(d) is low and shows no significant signs of slip bands, indicating that the identification was successful. Note that the rotation field and the residual fraction field is set to NaN (shown as grey in the figures) when the residual fraction is above its threshold or when the effective strain is below its threshold.

While this example may seem rather straightforward, the activities of the three Basal slip systems do not follow Schmid's law globally. In fact, the lowest SF Basal system, Basal-3 (SF of 0.07), shows most of the activity, while the highest SF system, Basal-1 (SF of 0.23), shows no activity. This could be due to the fact that the local loading conditions in the polycrystalline coating change the local stress state, rendering the SF insufficient to predict slip system activities. However, definite conclusions require measurements of the local stress state or advanced Transmission Electron Microscopy (TEM) analysis of dislocations, which is outside the scope of this work.

\subsection{Conclusive vs. Inconclusive Identification results}
\label{subSec:conclusivevsinconclusive}

\begin{figure}[htb!]
\centering
\includegraphics[width=0.85\textwidth]{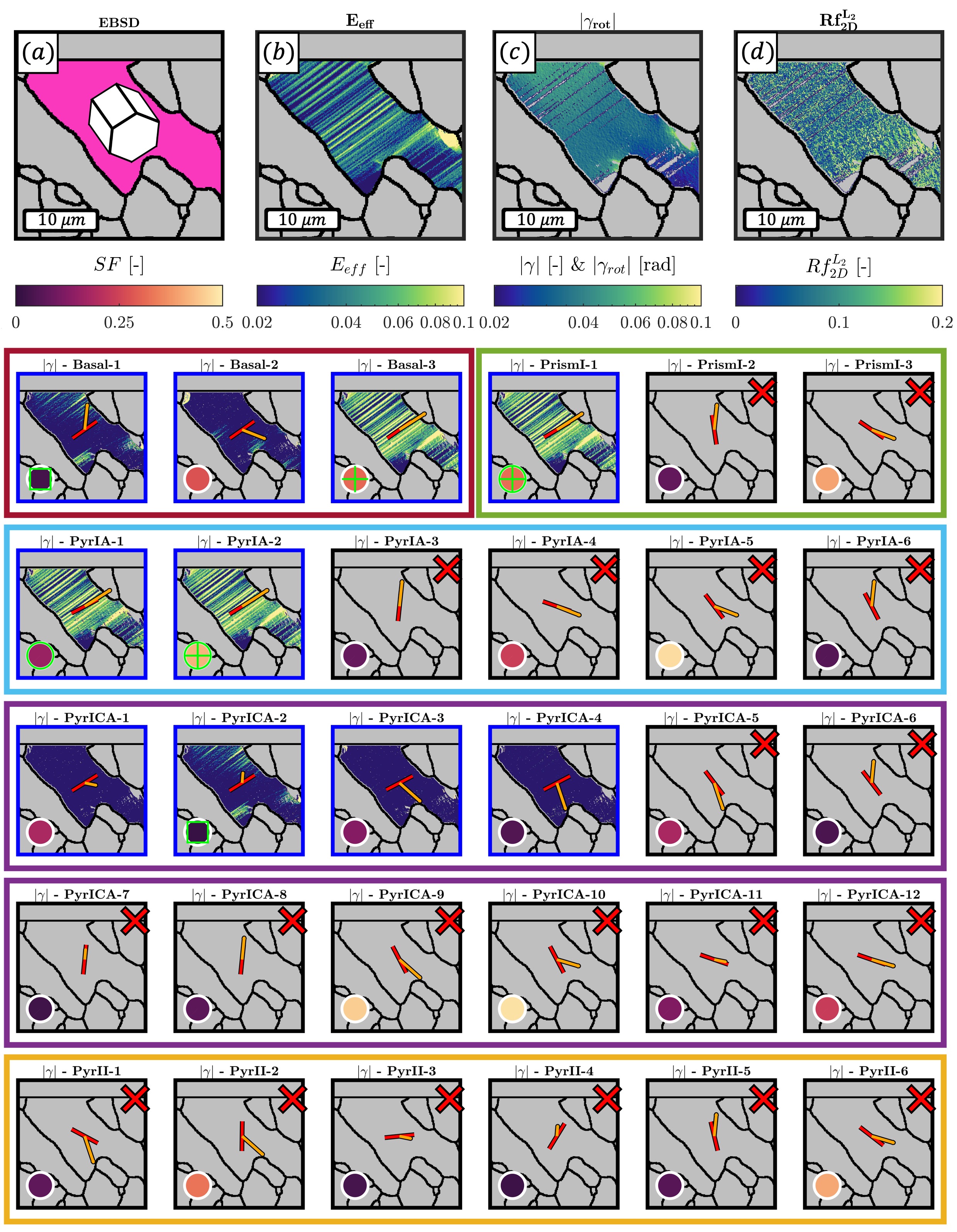}
\caption{Evaluation of slip system activities in an HCP Zn grain using the \psslip method, with a focus on nondiscriminatory slip systems. See the caption of \autoref{fig6} for the figure structure. Nondiscriminatory slip system groups are identified and marked with specific green symbols within the SF circles to denote their indistinguishability due to close angular proximity.}
\label{fig7}
\end{figure}

As explained in \autoref{subSec:inconcl}, the identification can be inconclusive when slip systems have equal 2D (in-plane) kinematics, of which \autoref{fig7} shows an example. The nondiscriminatory slip system groups, based on a 5 \degree angle threshold of the slip trace and projected slip direction, are annotated by specific green symbols within the SF circles of the slip system activity fields. Note that, in general, the in-plane similarity of slip systems can be judged visually by assessing the difference between the plotted slip trace and direction. In this case, there are 3 groups of nondiscriminatory systems. Slip systems PrismI-1 and PyrIA-2 are each part of two nondiscriminatory groups. The green rectangle shows that Basal-1 and PyrICA-2 are nondiscriminatory, but showing limited slip activities. The two other groups, annotated by the green "+" sign (Basal-3, PrismI-1 and PyrIA-2) and circles (PrismI-1, PyrIA-1 and PyrIA-2) are very close to each other, but different enough to be separated into two groups. These two groups have the largest slip activities and it is likely that only Basal-3 is actually active, since it has a significant SF ($SF>0.35$) and it is expected to have a low CRSS. Finally, this grain still has several unique slip systems, of which Basal-2 shows limited activity.

Another notable feature in this grain is the presence of rotation over almost the full grain, as shown in \autoref{fig7}(c). Considering that the magnitude of rotation (in radians) is similar to the slip magnitudes, regular SSLIP without a rotation-correction would likely fail here. The resulting rotation field is almost fully smooth, which one would expect for elastic rotation, and its presence over the full grain is a strong indication of grain rotation, which may be related to grain boundary sliding.  

\subsection{Non-Basal Multi-slip Activity in Zn}
\label{subSec:multiSlip}

\begin{figure}[htb!]
\centering
\includegraphics[width=0.85\textwidth]{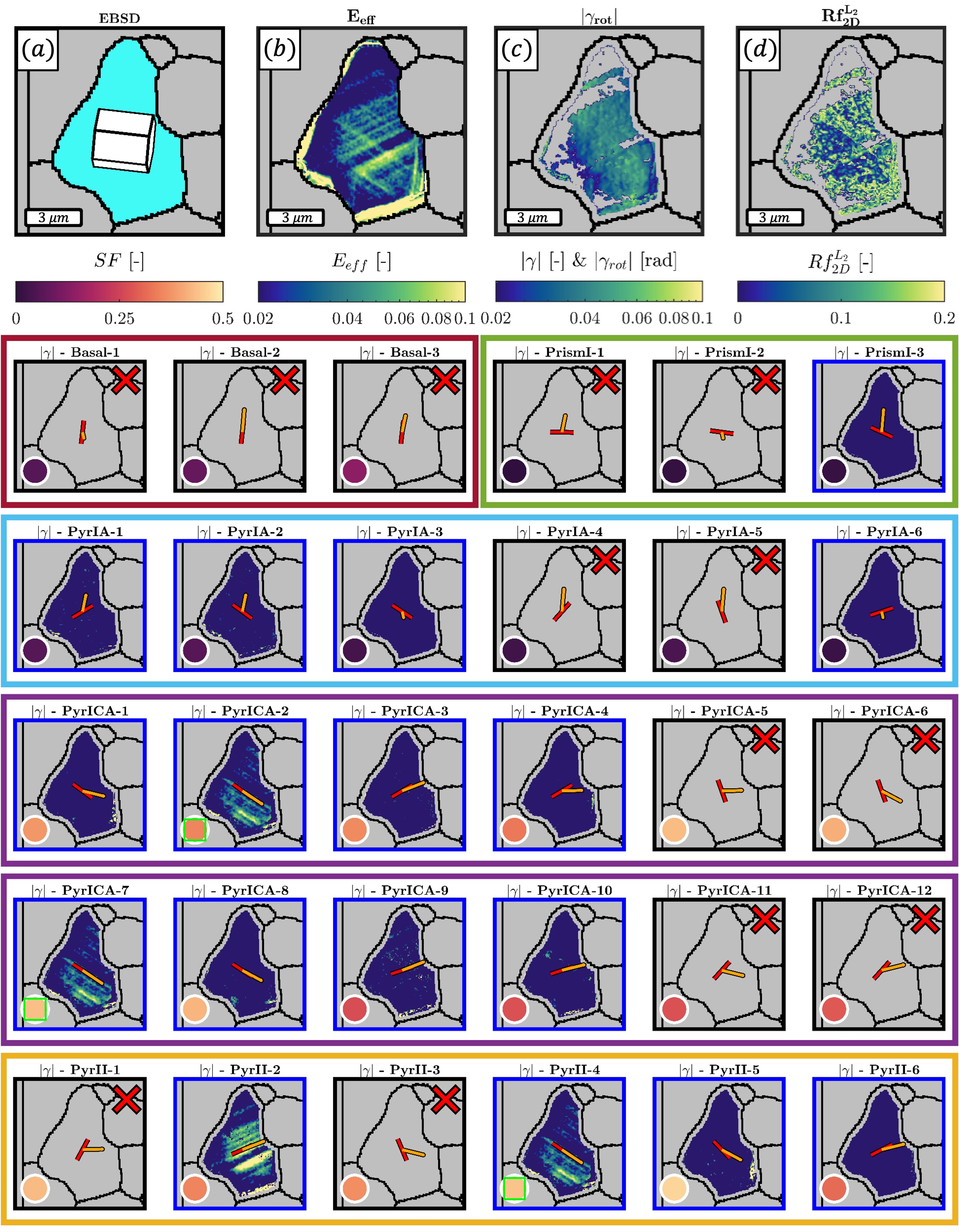}
\caption{Complex slip system interaction in a hexagonal close-packed Zinc grain analyzed by the \psslip method. See the caption of \autoref{fig6} for the figure structure.}
\label{fig8}
\end{figure}

After these first two examples with a single dominant slip trace in the strain map, we now turn to a more complex scenario where multiple slip systems with different trace orientations are simultaneously active. The Zinc grain analyzed in \autoref{fig8} shows a case where several slip systems are active. The deformation pattern as observed in the effective strain field (\autoref{fig8}(b)) reveals distinct arrays of slip bands, indicating the multi-slip nature of the deformation.

The \psslip method is applied here, including 17 slip systems after Radon transform preselection. While this number is significant, there is only 1 group of 3 nondiscriminatory slip systems: PyrICA-2, PyrICA-7 and PyrII-2. Notably, the Basal slip systems are not among those preselected.

The \psslip results shown in \autoref{fig8} clarify which slip systems are active. Slip system PyrII-2 is conclusively identified and the orientation of the slip bands conforms to the slip trace, which serves as an indirect validation of the results. The other part of the deformation is attributed to the nondiscriminatory group mentioned above, for which the traces also match the observed slip system activity field. Based on this, we can conclude that Pyramidal-II can occur in this Zn alloy, with potential activity also from Pyramidal-I CA. Additionally, it shows that \psslip can handle a large number of potentially active slip systems.

\subsection{Complex and Diffuse Slip Identified in Zn}
\label{subSec:complexSlip}

Now, we will push the limits of \psslip to identify plasticity in which slip traces are not trivially identified, predominantly because the slip is diffuse and spread out over the grain. While the original SSLIP method \cite{Vermeij2022AutomatedData} performed well in such cases (i.e. for FCC and BCC), the added preselection by the Radon transform for \psslip could prevent such a good performance here. Therefore, we evaluate the identification on two grains which show such complexity and diffuseness.

\begin{figure}[htb!]
\centering
\includegraphics[width=0.85\textwidth]{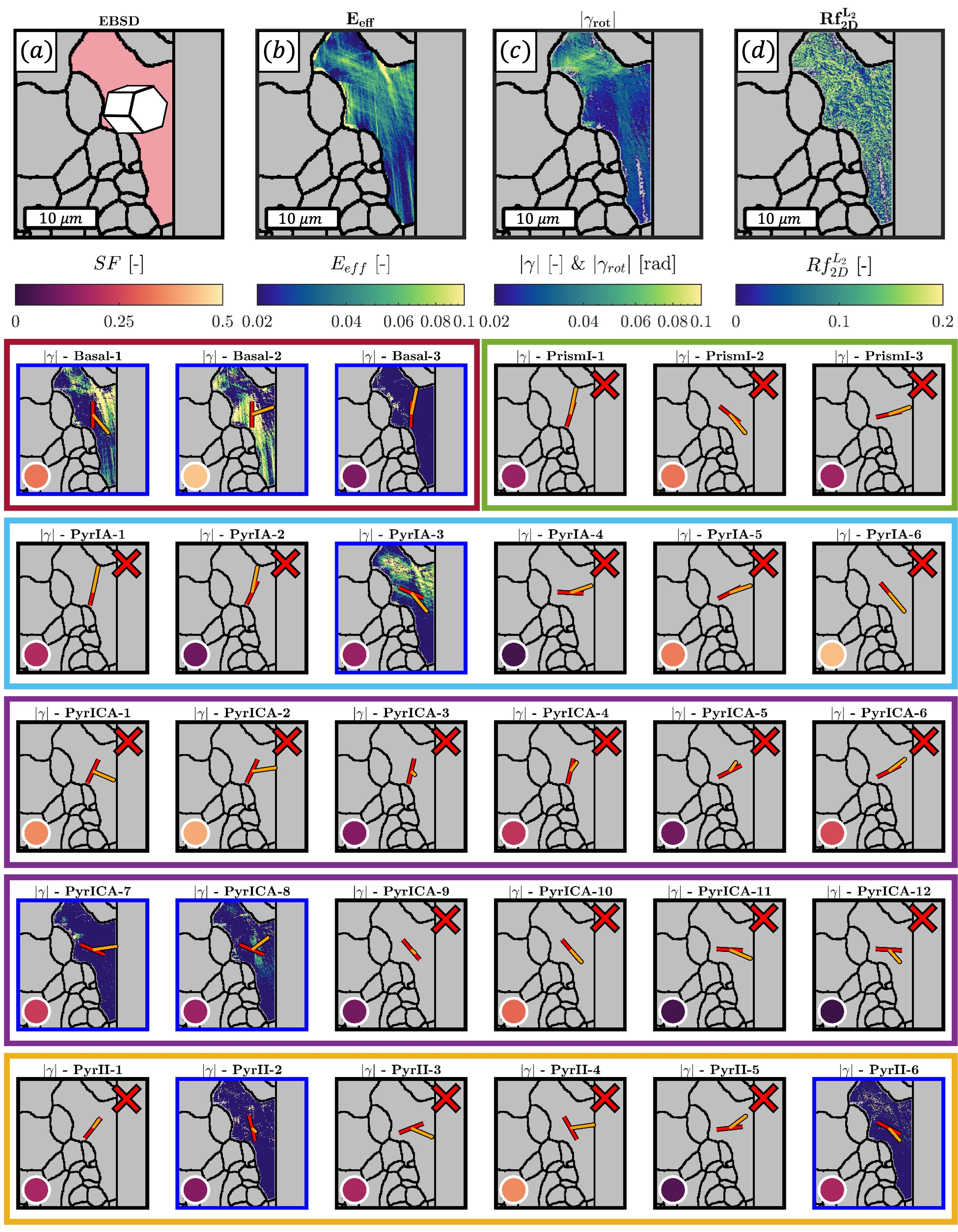}
\caption{Diffuse and complex plasticity, potentially cross slip, identified by \psslip. See the caption of \autoref{fig6} for the figure structure.}
\label{fig9}
\end{figure}

The grain depicted in \autoref{fig9} shows a complex deformation pattern, consisting of several sharp slip bands mixed with diffuse slip (\autoref{fig9}(b)). The Radon transform resolves 8 potential slip systems, all discriminatory, of which 3 systems show prominent slip activity as identified by \psslip. The activation of Basal-1 and particularly Basal-2 were expected due to their high SF and low expected CRSS, while the (near) absence of Basal-3 slip seems logical as it has the lowest SF of the three Basal systems. However, the Basal slip bands are less discrete and straight as compared to \autoref{fig6} and \autoref{fig7}. Additionally, PyrIA-3, which has a low SF (as is the case for all of the preselected non-Basal slip systems), shows high activity. While Pyramidal-IA has been observed before \cite{Parisot2004DeformationModes}, its occurrence is not common. PyrIA-3  and Basal-1 have the same slip direction (Burgers vector) and their activity fields show a certain degree of overlap. This indicates that cross slip may be active \cite{Vermeij2022AutomatedData}, although this was not apparent from our previous experiments on bi-crystal specimens \cite{Konig2024TheCoating}. Consequently, validation through advanced TEM characterization of dislocations would be required to draw firm conclusions.

\begin{figure}[htb!]
\centering
\includegraphics[width=0.85\textwidth]{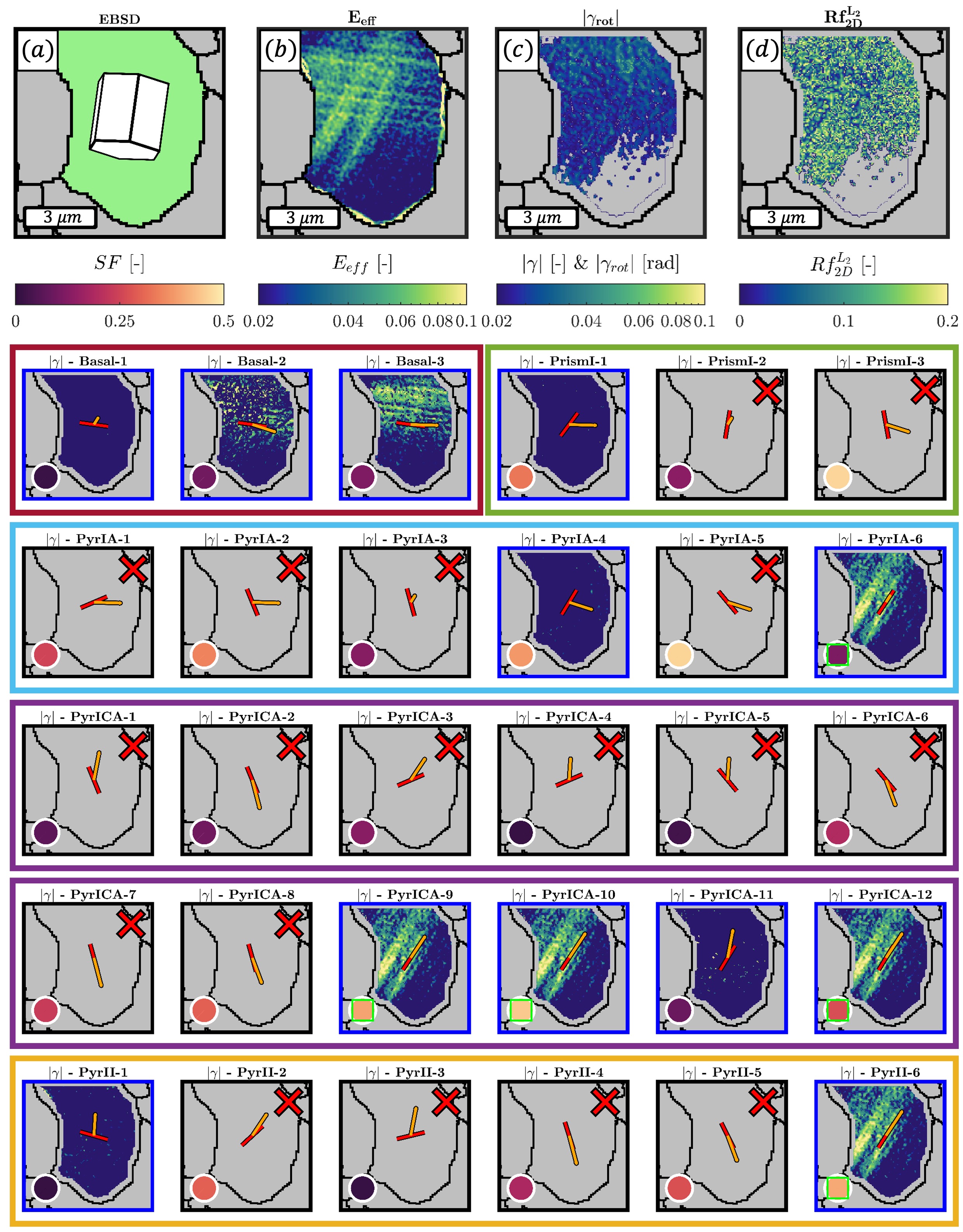}
\caption{Diffuse crystal plasticity identified by \psslip. See the caption of \autoref{fig6} for the figure structure.}
\label{fig10}
\end{figure}

Finally, we assess a grain with the highest level of complexity: a highly diffuse multi-slip deformation pattern in combination with a high number of Radon transform preselected slip systems, of which many are also nondiscriminatory, as depicted in \autoref{fig10}. Here the Radon transform results in 12 preselected slip systems, of which 5 are in a single nondiscriminatory group. Basal slip, with low SF, is conclusively identified. The differently oriented slip band activities are attributed to the nondiscriminatory group, which consists of PyrIA, PyrICA and PyrII slip systems. Notably, all slip activity fields show slip bands that concur with their respective slip trace orientations, again serving as an indirect validation. The diffuse deformation pattern is thereby appropriately unraveled, even for this highly complex case.  

\section{Conclusions}
\label{sec:concl}
In this work, we have proposed a significant extension of the original SSLIP (Slip System based Local Identification of Plasticity) method, termed as \psslip, addressing automated slip system identification on HCP metals, based on SEM-DIC and EBSD data. We have demonstrated that this methodology can effectively tackle the challenges posed by HCP crystal structures, where the presence of multiple slip families and the resulting high number of in-plane linearly dependent slip system kinematics can complicate the analysis.

Key advancements introduced in this work include (i) automated preselection of slip systems using the Radon transform, (ii) identification of the two best matching slip systems for each data point, (iii) the incorporation of a robust rotation correction mechanism, and (iv) a procedure to deal with slip systems with in-plane Schmid tensors that cannot be discriminated within the measurement error. This has allowed for a more precise and comprehensive identification of active slip systems, yielding the full slip system activity maps with all slip systems for each grain.

The performance of \psslip was first demonstrated successfully on a challenging virtual HCP case study on which the original SSLIP method did not perform well. Subsequently, we considered microstructure-correlated SEM-DIC deformation maps on a polycrystalline Zn anti-corrosion coating as an experimental case study, from which we have analyzed a number of challenging grains as proof of principle. It was found that \psslip overcomes the limitations of traditional trace-based analysis, particularly in cases where slip is diffuse or cross slip and diffuse slip are present. Additionally, where necessary, we provide a transparent overview of nondiscriminatory slip systems, which are fundamentally indistinguishable from only 2D (in-plane) deformation maps. 

The \psslip approach was confirmed to be an objective method that does not rely on factors such as the Schmid factor (SF) to select active slip systems. Every identified slip activity field shown in this paper can be indirectly validated by comparing its deformation pattern to the slip plane trace. \psslip offers a strong compromise between the original SSLIP method, in which preselection is not required, and SF based methods, especially for HCP metals since the critical resolved shear stress values for different slip families are often unknown. Additionally, the effect of neighbouring grains in a polycrystalline microstructure renders the SF estimates unreliable. 

In summary, the \psslip framework, of which the code is shared openly, stands as a significant contribution to the field of identification of crystal plasticity, offering a robust tool for investigating and understanding the plastic deformation behavior of HCP materials. The methodology paves the way for future studies to further explore and quantify the slip system activities under different loading conditions and in other HCP materials, which allows statistical studies and enhances the predictive capabilities of material models and simulations.

\section*{Acknowledgements}
We acknowledge Marc van Maris and Mark Vissers for experimental support.

This research was carried out as part of the “Next-Coat” project, under project number N19016b in the framework of the Partnership Program of the Materials innovation institute M2i (www.m2i.nl) and the Netherlands Organization for Scientific Research (http:// www.nwo.nl).

\section*{Code and Data Availability}
The Matlab code for the original SSLIP method and for the new \psslip method, with several examples, will be available on Github: \url{https://www.github.com/TijmenVermeij/SSLIP}. The full datasets are available upon request.

\centering
\noindent\rule{8cm}{0.4pt}



\begin{thebibliography}{10}

\bibitem{lunt2018quantification}
D.~Lunt, X.~Xu, T.~Busolo, J.~Quinta~da Fonseca, and M.~Preuss.
\newblock Quantification of strain localisation in a bimodal two-phase titanium
  alloy.
\newblock {\em Scripta Materialia}, 145:45--49, 2018.

\bibitem{githens2020characterizing}
A.~Githens, S.~Ganesan, Z.~Chen, J.~Allison, V.~Sundararaghavan, and S.~Daly.
\newblock {Characterizing microscale deformation mechanisms and macroscopic
  tensile properties of a high strength magnesium rare-earth alloy: A combined
  experimental and crystal plasticity approach}.
\newblock {\em Acta Materialia}, 186:77--94, 2020.

\bibitem{harte2020effect}
A.~Harte, M.~Atkinson, A.~Smith, C.~Drouven, S.~Zaefferer, J.~Quinta~da
  Fonseca, and M.~Preuss.
\newblock The effect of solid solution and gamma prime on the deformation modes
  in {Ni}-based superalloys.
\newblock {\em Acta Materialia}, 194:257--275, 2020.

\bibitem{vermeij2022nanomechanical}
T.~Vermeij, J.A.C. Verstijnen, T.J.J. Ramirez~y Cantador, B.~Blaysat,
  J.~Neggers, and J.P.M. Hoefnagels.
\newblock A nanomechanical testing framework yielding front\&rear-sided,
  high-resolution, microstructure-correlated {SEM-DIC} strain fields.
\newblock {\em Experimental Mechanics}, 62:1625--1646, 2022.

\bibitem{vermeij2023damage}
T.~Vermeij, C.J.A. Mornout, V.~Rezazadeh, and J.P.M. Hoefnagels.
\newblock Martensite plasticity and damage competition in dual-phase steel: A
  micromechanical experimental-numerical study.
\newblock {\em Acta Materialia}, 254:119020, 2023.

\bibitem{vermeij2023testing2d}
T.~Vermeij, J.~Wijnen, R.H.J. Peerlings, M.G.D. Geers, and J.P.M. Hoefnagels.
\newblock A quasi-2d integrated experimental--numerical approach to
  high-fidelity mechanical analysis of metallic microstructures.
\newblock {\em Acta Materialia}, 264:119551, 2024.

\bibitem{Parisot2004DeformationModes}
Rodolphe Parisot, Samuel Forest, Andr{\'e} Pineau, Fran{\c c}ois Grillon,
  Xavier Demonet, and Jean~Michel Mataigne.
\newblock {Deformation and damage mechanisms of zinc coatings on hot-dip
  galvanized steel sheets: Part I. Deformation modes}.
\newblock {\em Metallurgical and Materials Transactions A 2004 35:3},
  35(3):797--811, 2004.

\bibitem{bridier2005analysis}
F.~Bridier, P.~Villechaise, and J.~Mendez.
\newblock {Analysis of the different slip systems activated by tension in a
  $\alpha$/$\beta$ titanium alloy in relation with local crystallographic
  orientation}.
\newblock {\em Acta Materialia}, 53(3):555--567, 2005.

\bibitem{Bieler009}
T.R. Bieler, P.~Eisenlohr, F.~Roters, D.~Kumar, D.E. Mason, M.A. Crimp, and
  D.~Raabe.
\newblock {The role of heterogeneous deformation on damage nucleation at grain
  boundaries in single phase metals}.
\newblock {\em International Journal of Plasticity}, 25(9):1655--1683, 2009.

\bibitem{echlin2016incipient}
M.P. Echlin, J.C. Stinville, V.M. Miller, W.C. Lenthe, and T.M. Pollock.
\newblock {Incipient slip and long range plastic strain localization in
  microtextured {Ti-6Al-4V} titanium}.
\newblock {\em Acta Materialia}, 114:164--175, 2016.

\bibitem{OROZCOCABALLERO2017367}
A.~Orozco-Caballero, D.~Lunt, J.D. Robson, and J.~{Quinta da Fonseca}.
\newblock {How magnesium accommodates local deformation incompatibility: A
  high-resolution digital image correlation study}.
\newblock {\em Acta Materialia}, 133:367--379, 2017.

\bibitem{harr2021effect}
M.E. Harr, S.~Daly, and A.L. Pilchak.
\newblock {The effect of temperature on slip in microtextured
  {Ti-6Al-2Sn-4Zr-2Mo} under dwell fatigue}.
\newblock {\em International Journal of Fatigue}, 147:106173, 2021.

\bibitem{Hu2023AutomatedAlloy}
H.~Hu, F.~Briffod, T.~Shiraiwa, and M.~Enoki.
\newblock {Automated slip system identification and strain analysis framework
  using high-resolution digital image correlation data: Application to a
  bimodal Ti-6Al-4V alloy}.
\newblock {\em International Journal of Plasticity}, 166:103618, 7 2023.

\bibitem{sperry2020slip}
R.~Sperry, A.~Harte, J.~Quinta~da Fonseca, E.R. Homer, R.H. Wagoner, and D.T.
  Fullwood.
\newblock {Slip band characteristics in the presence of grain boundaries in
  nickel-based superalloy}.
\newblock {\em Acta Materialia}, 193:229--238, 2020.

\bibitem{chen_daly_2016}
Z.~Chen and S.H. Daly.
\newblock Active slip system identification in polycrystalline metals by
  {Digital Image Correlation (DIC)}.
\newblock {\em Experimental Mechanics}, 57(1):115--127, 2016.

\bibitem{BOURDIN2018307}
F.~Bourdin, J.C. Stinville, M.P. Echlin, P.G. Callahan, W.C. Lenthe, C.J.
  Torbet, D.~Texier, F.~Bridier, J.~Cormier, P.~Villechaise, T.M. Pollock, and
  V.~Valle.
\newblock Measurements of plastic localization by heaviside-digital image
  correlation.
\newblock {\em Acta Materialia}, 157:307 -- 325, 2018.

\bibitem{XU2019376}
X.~Xu, D.~Lunt, R.~Thomas, R.~Prasath Babu, A.~Harte, M.~Atkinson, J.~Quinta
  da~Fonseca, and M.~Preuss.
\newblock Identification of active slip mode in a hexagonal material by
  correlative scanning electron microscopy.
\newblock {\em Acta Materialia}, 175:376 -- 393, 2019.

\bibitem{STINVILLE2020172}
J.C. Stinville, P.G. Callahan, M.A. Charpagne, M.P. Echlin, V.~Valle, and T.M.
  Pollock.
\newblock Direct measurements of slip irreversibility in a nickel-based
  superalloy using high resolution digital image correlation.
\newblock {\em Acta Materialia}, 186:172 -- 189, 2020.

\bibitem{sperry2021comparison}
R.~Sperry, S.~Han, Z.~Chen, S.H. Daly, M.A. Crimp, and D.T. Fullwood.
\newblock Comparison of {EBSD, DIC, AFM, and ECCI} for active slip system
  identification in deformed {Ti-7Al}.
\newblock {\em Materials Characterization}, 173:110941, 2021.

\bibitem{poole2024high}
B.~Poole, A.~Marsh, D.~Lunt, C.~Hardie, M.~Gorley, C.~Hamelin, and A.~Harte.
\newblock High-resolution strain mapping in a thermionic lab6 scanning electron
  microscope.
\newblock {\em Strain}, 2024.

\bibitem{black2024micro}
R.L. Black, D.~Anjaria, J.~Gen{\'e}e, V.~Valle, and J.C. Stinville.
\newblock Micro-strain and cyclic slip accumulation in a polycrystalline
  nickel-based superalloy.
\newblock {\em Acta Materialia}, 266:119657, 2024.

\bibitem{Vermeij2022AutomatedData}
T.~Vermeij, R.H.J. Peerlings, M.G.D. Geers, and J.P.M. Hoefnagels.
\newblock {Automated Identification of Slip System Activity Fields from Digital
  Image Correlation Data}.
\newblock {\em Acta Materialia}, 243:118502, 1 2022.

\bibitem{stinville2023insights}
J.C. Stinville, M.A. Charpagne, R.~Maa{\ss}, H.~Proudhon, W.~Ludwig, P.G.
  Callahan, F.~Wang, I.J. Beyerlein, M.P. Echlin, and T.M. Pollock.
\newblock {Insights into Plastic Localization by Crystallographic Slip from
  Emerging Experimental and Numerical Approaches}.
\newblock {\em Annual Review of Materials Research}, 53:275--317, 2023.

\bibitem{wijnen2024high}
J.~Wijnen, T.~Vermeij, J.P.M. Hoefnagels, M.G.D. Geers, and R.H.J. Peerlings.
\newblock High-resolution numerical-experimental comparison of heterogeneous
  slip activity in quasi-2d ferrite sheets.
\newblock {\em arXiv preprint arXiv:2402.16199}, 2024.

\bibitem{scherertensile}
J.~Scherer, J.~Hure, R.~Madec, F.~Le~Bourdais, L.~van Brutzel, S.~Sao-Joao,
  G.~Kermouche, J.~Besson, and B.~Tanguy.
\newblock {Tensile and micro-compression behaviour of AISI 316L austenitic
  stainless steel single crystals at 20$\,^{\circ}$ C and 300$\,^{\circ}$ C:
  experiments, modeling and simulation}.
\newblock {\em Materials Science and Engineering: A}, page 146471, 2024.

\bibitem{agius2023experimental}
D.~Agius, D.~Cram, C.~Hutchinson, M.~Preuss, Z.~Sterjovski, and C.~Wallbrink.
\newblock {An experimental and computational study into strain localisation in
  beta-annealed Ti-6Al-4V}.
\newblock {\em Procedia Structural Integrity}, 45:4--11, 2023.

\bibitem{depriester2023crystal}
D.~Depriester, J.P. Goulmy, and L.~Barrallier.
\newblock Crystal plasticity simulations of in situ tensile tests: A two-step
  inverse method for identification of cp parameters, and assessment of cpfem
  capabilities.
\newblock {\em International Journal of Plasticity}, 168:103695, 2023.

\bibitem{yin2023three}
W.~Yin, F.~Briffod, H.~Hu, T.~Shiraiwa, and M.~Enoki.
\newblock Three-dimensional configuration of crystal plasticity in stainless
  steel assessed by high resolution digital image correlation and confocal
  microscopy.
\newblock {\em International Journal of Plasticity}, 170:103762, 2023.

\bibitem{Konig2024TheCoating}
D.~Konig, T.~Vermeij, F.~Maresca, and J.P.M. Hoefnagels.
\newblock Direct comparison of nanoscale plasticity in single and bi-crystal
  tensile tests extracted from a zinc coating.
\newblock {\em Materials Science and Engineering: A}, 932:148128, 2025.

\bibitem{helgason1999radon}
S.~Helgason.
\newblock {\em The radon transform}, volume~2.
\newblock Springer, 1999.

\bibitem{Paupler1988G.0070168938}
P.~Paupler.
\newblock {\em {G. E. Dieter. Mechanical Metallurgy. 3rd ed., Mc Graw-Hill Book
  Co., New York 1986. XXIII + 751 p., DM 138.50, ISBN 0--07--016893--8}},
  volume~23.
\newblock 1988.

\bibitem{peaks2}
Kristupas Tikui{\v s}is.
\newblock {peaks2 - find peaks in 2D data without additional toolbox
  (https://www.mathworks.com/matlabcentral/fileexchange/113225-peaks2-find-peaks-in-2d-data-without-additional-toolbox),
  MATLAB Central File Exchange.}

\bibitem{LENTHE2019112841}
W.C. Lenthe, S.~Singh, and M.~De Graef.
\newblock A spherical harmonic transform approach to the indexing of electron
  back-scattered diffraction patterns.
\newblock {\em Ultramicroscopy}, 207:112841, 2019.

\bibitem{HoefnagelsPattern}
J.P.M. Hoefnagels, M.P.F.H.L. van Maris, and T.~Vermeij.
\newblock One-step deposition of nano-to-micron-scalable, high-quality digital
  image correlation patterns for high-strain in-situ multi-microscopy testing.
\newblock {\em Strain}, 55(6):e12330, 2019.
\newblock e12330 STRAIN-1507.R1.

\bibitem{Vermeij2021}
T.~Vermeij and J.P.M. Hoefnagels.
\newblock Plasticity, localization, and damage in ferritic-pearlitic steel
  studied by nanoscale digital image correlation.
\newblock {\em Scripta Materialia}, 208, 2021.

\bibitem{Mornout2024LIBS}
C.J.A. Mornout, G.~Slokker, T.~Vermeij, D.~Konig, and J.P.M. Hoefnagels.
\newblock Slide: Automated identification and quantification of grain boundary
  sliding and opening in 3d.
\newblock {\em arXiv:2504.04898}, 2025.

\bibitem{bednarczyk2023determination}
W.~Bednarczyk, M.~Watroba, M.~Jain, K.~Mech, P.~Bazarnik, P.~Ba{\l}a,
  J.~Michler, and K.~Wieczerzak.
\newblock {Determination of critical resolved shear stresses associated with
  $<a>$ slips in pure Zn and Zn-Ag alloys via micro-pillar compression}.
\newblock {\em Materials \& Design}, 229:111897, 2023.

\end{thebibliography}

\end{document}